\documentclass[aps,preprint,prd,aps,12pt]{revtex4-2}
\usepackage{graphicx}
\usepackage[english]{babel}
\usepackage[T1]{fontenc}
\usepackage{mathrsfs}
\usepackage[tbtags]{amsmath}
\usepackage{amssymb}
\usepackage{amsxtra}
\usepackage{amsopn}
\usepackage{latexsym}
\usepackage[mathcal]{eucal}
\usepackage{mathtools}

\newcommand{\BE}{\begin{equation}}
\newcommand{\EE}{\end{equation}}
\newcommand{\BA}{\begin{align}}
\newcommand{\EA}{\end{align}}

\newcommand{\nn}{\nonumber}

\newcommand{\ppp}{ \frac{{\rm d}^4p}{(2\pi)^4}}

\newcommand{\Rerm}{\mathop{\rm Re}}

\renewcommand{\Im}{\mathop{\rm Im}}

\newcommand{\unictandinfnct}{Dipartimento di Fisica e Astronomia dell'Universit\`a di Catania, INFN Sezione di Catania, Via S.Sofia 64, I-95123 Catania, Italy}

\begin{document}

\title{Analytic continuation and physical content of the gluon propagator}

\author{Fabio Siringo}\email{fabio.siringo@ct.infn.it}\affiliation{\unictandinfnct}
\author{Giorgio Comitini}\email{giorgio.comitini@dfa.unict.it}\affiliation{\unictandinfnct}

\date{\today}

\begin{abstract}
The analytic continuation of the gluon propagator is revised in the light of recent findings on the possible existence of complex conjugated poles. The contribution of the anomalous pole must be added when Wick rotating, leading to an effective
Minkowskian propagator which is not given by the trivial analytic continuation of the Euclidean function.
The effective propagator has an integral representation in terms of a spectral function which is naturally related to a set of elementary (complex) eigenvalues of the Hamiltonian, thus generalizing the usual K\"all\'en-Lehmann description. 
A simple toy model shows how the elementary eigenvalues might be related to actual physical quasiparticles of the
non-perturbative vacuum.
\end{abstract}




\maketitle 
\newpage

\section{Introduction}

The gluon and quark propagators play a very important role in the study of strong interactions and a detailed knowledge
of the real-time correlators would provide the basic blocks for a study of heavy-ion collisions from first principles.
However, in the low energy nonperturbative regime of strong interactions, our knowledge of the propagators is
very limited and usually based on numerical calculations 
in the Euclidean space, including lattice simulations\cite{olive,aguilar04,cucch08,cucch08b,cucch09b,bogolubsky,
olive09,dudal,binosi12,olive12,burgio15,duarte}  
and continuum studies\cite{papa15b,aguilar8,aguilar10,aguilar14,papa15,fischer2009,huber14,huber15g,
huber15b,pawlowski10,pawlowski10b,pawlowski13,
varqcd,genself,watson10,watson12,rojas,var,qed,higher,reinhardt04,reinhardt05,reinhardt14}.
Thus, the problem of analytic
continuation from Euclidean to Minkowski space 
is still under intense debate\cite{dudal14,haas,qin,christ,cyrol,roth,cyrol,horak,binosi,dudal20,dispersion,kondo18,kondo21}. 

For a generic field theory which describes
physical particles, many exact results have been developed in the past
and some of them have  been even
extended to N-point functions\cite{abrikosov,fetter,kapusta,evans,cuniberti}.  
If we did not know about confinement, then the non-Abelian gauge theory would 
be expected to satisfy the same general conditions which hold for all physical particles: the propagators should
be characterized by the usual analytic properties and could be written by the standard K\"all\'en-Lehmann
integral representation in terms of a positive defined spectral function. Then, the knowledge of the spectral function
would allow a trivial analytic continuation from Euclidean to Minkowski space. Actually, from the formal point of view,
there is nothing in the Lagrangian which might foreshadow a different behavior for the correlators of QCD in comparison to, for instance, QED. For the same reason, we still miss a full understanding of confinement. On the other hand, the 
interacting compact QED seems to follow the same anomalous features of Yang-Mills theory\cite{orlando22}.

Because of color confinement, gluon and quarks are usually regarded as internal degrees of freedom of the theory.
More precisely, they do not occur in the asymptotic states, but they do exist as quasiparticles in a very hot
quark-gluon plasma above the deconfinement transition. Thus, they cannot be regarded as totally unphysical mathematical
degrees of freedom like a ghost. But, since we cannot detect a free gluon or a free quark, some 
unitarity constraints might be relaxed for these particles and there is no reason to believe that the same positivity
conditions should still hold for their spectral functions. Moreover, we don't have any formal proof that there is any
spectral representation at all, so that  the usual analytic properties of the propagators might be questioned. 
That explains why the problem of analytic continuation is still so strongly debated.

On the other hand, we believe by now that QCD is a complete consistent theory 
which generates its IR cutoff dynamically
and we expect that the confinement must arise from the same Lagrangian, as it actually happens in the lattice, 
without adding spurious effects by hand. Thus, it is also very reasonable to expect that the exact propagators of the
theory should be substantially different than the other propagators of the standard model. Somehow, some sign of
confinement must appear dynamically in the structure of the propagators and must be buried in their analytic
properties, in the complex plane. But, since our most accurate information on the propagators 
is found numerically in the Euclidean space, we have no direct knowledge of the analytic properties in the complex plane,
and the continuation has the nature of a guessing work.
Moreover, there are many clues that the analytic structure is untrivial. For instance, from lattice 
and continuous calculations we know
that the curvature of the propagator changes sign and the Schwinger function crosses the zero, becoming 
negative at a length of some Fermi units\cite{alkofer04,xigauge}. 
These are all signs of a spectral function which is not positive defined, if there is a spectral function. What is even more disturbing, there are independent predictions of complex conjugated poles which invalidate the K\"all\'en-Lehmann spectral representation, even if the spectral density were negative\cite{dispersion,kondo18}.
Complex poles were predicted by effective models\cite{GZ,stingl,dudal18} for the gluon propagator in the past.
From first principles, their existence arises from a one-loop screened 
expansion of the exact Lagrangian\cite{ptqcd,ptqcd2,analyt,xigauge,damp,therm,varT,RG,beta,ghost}. 
But they also occur in one-loop approximations\cite{kondo18} of effective models 
like Curci-Ferrari\cite{tissier10,tissier11,tissier14,serreau,reinosa,pelaez,pelaez21,2LQ}.

Many numerical attempts
at reconstructing the gluon propagator and its spectral function have shown a better agreement with the data if
a pole part, with complex conjugated poles, is added to the usual spectral integral\cite{binosi,wink}.
Even the outcome of Schwinger-Dyson euqations in the complex plane seems to suggest the existence of
singularities outside the real axis\cite{fischer}. The quark propagator has also been 
reported to show complex conjugated poles by the one-loop screened expansion\cite{quark}
 and a general study of the pole structure in one-loop approximations has been discussed in Ref.\cite{kondo18}.

While there are reconstruction methods which describe the lattice data without requiring the existence of
complex poles\cite{cyrol,horak}, the quality of the reconstruction seems to improve when the poles are added\cite{wink}.
Then the issue of the existence and dynamical meaning of the complex poles becomes of paramount importance.

In this paper, we discuss how a consistent quantum theory can be recovered when there are complex poles in the Euclidean propagator. Assuming that complex conjugated poles do exist in the exact propagator and that they might play
a physical role in the confinement mechanism\cite{excited}, we show how well defined propagators can be actually derived
in real time by a modified Wick rotation. 
Then, we see how a modified  K\"all\'en-Lehmann spectral representation, including the anomalous 
pole part, can be derived from first principles in presence of zero-norm states, with complex energies. Finally
we speculate on a direct relation between the complex energies and a set of observable glueball physical states.

While we cannot say if the complex poles are genuine and if they do exist at all in the exact gluon propagator, 
here we show how their
existence would lead to untrivial consequences in the analytic continuation to real time. 

The paper is organized as follows: the problem of analytic continuation is discussed in Sec.~II and
the standard Wick rotation is recovered in Sec.~III in order to fix the notation; in Sec.~IV
a modified analytic continuation is derived by two different methods, 
by residue subtraction and by convergence arguments, yielding an
interesting spectral representation; in Sec.~V the same  anomalous spectral density is derived from first principles
as a modified K\"all\'en-Lehmann representation in presence of complex eigenvalues; in Sec.~VI a Hermitian  toy  model
is discussed which leads to a speculative physical interpretation of the anomalous spectral density in terms of physical states; finally, in Sec.~VII, the main results are summarized and discussed.

\section{Analytic continuation of the gluon propagator}

While most of the rigorous results in quantum field theory have been established in the Euclidean space, 
the physical content of a theory
is usually extracted in Minkowski space. However, if there are complex conjugated poles, 
a general rigorous connection between amplitudes in Euclidean and  Minkowski spaces is missing
because the singularities do not allow the usual Wick rotation
and the standard K\"all\'en-Lehmann spectral representation does not hold\cite{dispersion,kondo18,kondo21}.
Thus, the extraction of the physical content from the theory 
might be quite tricky and might rely on some guessing work.
Moreover, 
the numerical knowledge of an amplitude on the real axis of the Euclidean space is usually not enough
for reconstructing its analytic continuation to Minkowski space\cite{cyrol,horak,binosi,dudal20,cuniberti}.

In perturbation theory, it is assumed and generally found that the Fourier Transform (F.T.) of the physical amplitudes have poles in the
second and fourth quadrant of the complex-energy plane and a branch cut on the real axis. Then, Wick rotation is allowed and gives
a well defined connection between the physics which occurs in Minkowski space and the amplitudes which are evaluated in the Euclidean space.
In that case, we find a circular path going: (i) from real time to real energy (by a F.T.); (ii) to the Euclidean space 
through Wick rotation in the complex-energy plane; (iii) to imaginary
time by an inverse F.T.; and as shown in the following line,
\BE
t \>\overset{\text{\,F.T.}} \Longleftrightarrow \> p_0 \> \overset{\text{\,Wick}} \Longleftrightarrow \> ip_4 \>\overset{\text{\,F.T.}} \Longleftrightarrow \>-i\, x_4 \>\overset{\text{$\tau$ order}} \Longleftrightarrow t
\label{circle}
\EE
(iv) the circle closes if a well defined prescription is given for the analytic continuation from imaginary time to real time. Time-ordered functions are not analytic in time, because of the functions $\theta(\pm t)$, then the relation between real-time and imaginary-time is not unique, in principle. 
The position $t=-i\tau$, where $t=x^0$ in Minkowski space and $\tau=x^4$ in the Euclidean space, can be {\it explained} by the physical motivation of mapping the time-evolution operator $U(t)=\exp(-i H t)$ on a thermal average by $U(-i\tau)=\exp(-\tau H)$ where
$0\le\tau\le \beta$. If we look at the general structure of a time-ordered correlator
\BE
\langle 0\vert T[A(t)B(0) \vert 0\rangle=\theta(t)\sum_n \rho_n\>  e^{-iE_nt}
+\theta(-t)\sum_n \rho^\prime_n \> e^{iE_nt}
\label{genT}
\EE
we find positive frequencies for $t>0$ and negative frequencies for $t<0$, which can be seen as antiparticle states going backwards in time. The position $t=-i\tau$ gives a weight $\exp(-E_n\tau)$ for positive frequencies and a weight
$\exp(E_n\tau)$ for negative frequencies. The correct thermal weight $\exp(-E_n \vert \tau\vert)$ is obtained in all cases if $\tau<0$ when $t<0$ and vice versa.
Thus, we generally {\it assume} that the generic time-ordered average transforms according to
\BE
\theta (t)\langle A(t)B(0) \rangle\>+\>
\theta (-t)\langle  B(0)A(t) \rangle \Longrightarrow 
\theta (\tau)\langle A(-i\tau)B(0) \rangle
\>+\>\theta (-\tau)\langle  B(0)A(-i\tau)\rangle
\EE
when going to the Euclidean space. Then, if the functions $\langle A(t)B(0)\rangle$ are analytic functions, there is a well defined and unique way to connect real-time amplitudes and imaginary-time averages.  
With the imaginary-time order 
understood in the analytic continuation, the circle is closed and we have a well defined connection among the different representations of the same theory as shown in Eq.(\ref{circle}). 
For a physical particle, which is present in the asymptotic
states, causality and unitarity determine the K\"all\'en-Lehmann spectral representation, giving a formal proof of the relation between time order and imaginary-time order. 
Thus our physical motivation is based on a solid formal  background\cite{abrikosov,fetter,kapusta,evans}.

While everything works fine in perturbation theory, in non-perturbative studies
the analytic properties of the amplitudes might not allow the usual Wick rotation. 
It happens for the gluon propagator which has been reported 
to show pairs of complex-conjugated poles by very different 
approaches\cite{binosi,GZ,stingl,wink,ptqcd,ptqcd2,analyt,xigauge,damp,therm,kondo18,fischer}.
Moreover,
even the analytic continuation from real time to imaginary time can be questioned on general grounds.
The generic time-ordered average in Eq.(\ref{genT}) might contain different parts which can be written as
\BE
\theta (t)\langle 0\vert A(t)B(0) \vert 0\rangle=\theta (t)\langle 0\vert A_1(t)B_1(0) \vert 0\rangle
+\theta (t)\langle 0\vert A_2(t)B_2(0) \vert 0\rangle\cdots
\EE
Assuming that the single averages on the right hand side are analytic functions of time, the analytic continuation would give
\BE
t=-i\tau\> \Longrightarrow \>\theta (\pm\tau)\langle 0\vert A_1(t)B_1(0) \vert 0\rangle
+\theta(\pm\tau)\langle 0\vert A_2(t)B_2(0) \vert 0\rangle\cdots
\label{thetapm}
\EE
where, in principle, each $\pm$ sign can depend on the properties of the specific operators in the average.
Some anomalous $\theta (-\tau)$ function, with the wrong sign, could be present in anomalous terms which might arise 
from an untrivial vacuum structure. For instance, when states with negative norm are present, we might find a superposition of zero-norm complex-conjugated states in the vacuum. The existence of eigenstates with a complex energy $E=-\omega-i\eta$,
with $\omega,\eta>0$, only makes sense if $t>0$ and $\tau<0$ since 
$\exp(-iEt)= \exp(-\eta t)\exp(i\omega t)=\exp(\omega\tau)\exp(i\eta \tau)$.
Thus, the existence of complex-conjugated poles might jeopardize the plain analytic continuation to imaginary time, with
a correspondence between time ordering and imaginary-time ordering which depends on the behavior of the single operators.
On the other hand, from a formal point of view, the K\"all\'en-Lehmann spectral representation is not valid in these anomalous cases\cite{dispersion}, and no general prescription is known for the analytic continuation.

The circle in Eq.(\ref{circle}) would be broken in two points:

\BE
t \>\overset{\text{\,F.T.}} \Longleftrightarrow \> p_0 \> \overset{\text{\, ?}} \longleftrightarrow \> ip_4 \>\overset{\text{\,F.T.}} \Longleftrightarrow \>-i\, x_4 \>\overset{\text{ ? }} \longleftrightarrow t
\label{brokencircle}
\EE

We end up with two different theories: one of them is defined in real-time Minkowski space, 
the other one in imaginary-time Euclidean space. Thus, it is not obvious what the physical content of the Euclidean theory is.

A perturbative-minded approach would be to assume that the plain analytic continuation can be used from real to imaginary energy even when Wick rotation is not allowed: i.e. assume that the amplitude $\Delta_E(p_4)$ in the Euclidean space is related to the amplitude in Minkowski space $\Delta_M(p_0)$ by the same analytic continuation
\BE
\Delta_E(p_4)=\Delta_M(ip_4)
\label{plain}
\EE
which holds in perturbation theory (as dicussed by Stingl\cite{stingl}). For instance, 
in the screened massive expansion\cite{ptqcd,ptqcd2,analyt,xigauge,damp,therm,varT,RG,beta,ghost,SD}, 
plain perturbation theory is used
for evaluating a gluon propagator which turns out to have complex-conjugated poles in all quadrants of the 
complex-energy plane.
The same expansion can be developed in the Euclidean or in the Minkowskian formalism yielding the same identical results up to the analytic continuation of Eq.(\ref{plain}).
However, since the Wick rotation is not allowed, the physical content in the two formalisms would be different. If the propagator is integrated together with other functions, in the calculation of some observable quantities, the result would be different in Euclidean and Minkowski space, because the Wick rotation would encounter the ``wrong'' poles, adding new contributions from the residues. By the same argument, going to real time on one side of Eq.(\ref{brokencircle}) and to imaginary time on the other side, the resulting amplitudes 
would  not be related by any analytic continuation in time. Yet, we could just assume that the usual analytic continuation is not valid in the direct space. In fact, by this approach, the F.T. of the gluon propagator gives reasonable results
even when complex poles are present. The two-point correlator turns out to be exponentially damped in imaginary time
(Schwinger function) and in the real time (propagator) when the energy is integrated by Jordan lemma and the contribution of the complex residues is correctly taken. It is quite obvious that the two functions are not related by the usual analytic continuation in time, which would transform an oscillating function in a divergent function. In principle, there is nothing wrong since the dynamics of a physical system in the Minkowski space might be different from the imaginary-time behavior of the corresponding Euclidean system. But we have still two different theories, depending on the space where they are defined.

According to a more formal approach to quantum field theory, the physical content of the theory should be
reconstructed starting from the Euclidean formalism. As discussed in Ref.\cite{kondo21}, one could assume, {\it as a starting point}, 
that the time ordered amplitudes are the analytic continuation of the imaginary-time amplitudes according to the standard
ordering in imaginary time
\BE
\langle 0\vert T\left\{A(t_1) B(t_2)\cdots \right\}\vert 0\rangle=\langle 0\vert T_\tau\left\{A(-i\tau_1) B(-i\tau_2)\cdots \right\}\vert 0\rangle
\label{plaint}
\EE
where $t=-i\tau$ and $T_\tau$ denotes an ordering in the imaginary-time $\tau$.
Having closed the chain on the right hand side of Eq.(\ref{brokencircle}) we can determine a unique way for connecting the Fourier transforms going
through the direct space. It turns out that the Euclidean and Minkowskian amplitudes are not connected by a plain analytic continuation in the energy
plane, which is the point where the circle breaks. The analysis of Ref.\cite{kondo21} leads to an unphysical gluon with a diverging correlator in
the real time. As expected, the oscillating Schwinger function gives diverging exponentials in real time, for $t\to\infty$ 
and $t\to -\infty$.
The complex-conjugated poles are then considered as unphysical features of a gluon state which does not belong to the physical Hilbert
space. Of course, if we started from the Minkowski space we would find a reasonable damped propagator (as discussed by Stingl\cite{stingl}) and an unphysical Schwinger function in the Euclidean space by the standard analytic continuation to imaginary time.
The result is unsatisfactory for several reasons, as it looks like we threw the baby out with the bath water.
Since the gluon has real effects in the phenomenology, giving rise to real physical jets and quasiparticles in the hot matter, 
it is not satisfactory that
the confinement might be explained by downgrading the gluon to a totally unphysical degree of freedom of the theory. 
Moreover, some mathematical degrees of freedom which do not exist in the real world, like the longitudinal photon or a ghost,
are not confined by any dynamical mechanism. On the other hand, the complex poles and residues of the gluon propagator seem to be even gauge-parameter-independent\cite{xigauge,nielsen}, 
pointing to some physical role of the pole part of the propagator\cite{excited}. 
We also mention that complex-conjugated poles have been found in the propagator of  the quark\cite{quark}, which is another physical (confined) particle.

We observe that Eq.(\ref{plaint}) is  the opposite assumption of Eq.(\ref{plain}), but neither of them 
might be valid in a general context.
Here, we would like to explore a third assumption, physically motivated, which can be regarded as an improved version of the perturbative-minded
plain continuation in the energy plane of Eq.(\ref{plain}). Assuming that the physical content should be reconstructed starting from the
Euclidean formalism, we look for a connection of the chain in the complex-energy plane in order to leave unchanged the content of the theory even
when the Wick rotation cannot be used. It turns out that when complex-conjugated poles are present, the analytic continuation of Eq.(\ref{plain}) must be
supplemented by adding the residues of the poles which are encountered by the Wick rotation. The same procedure was used in Ref.\cite{detmold},  and
found successful for extracting the physical content of the amplitude when a pole does not allow the usual Wick rotation.
The method ensures that the physical content of the theory does not change when going from the Euclidean to the Minkowskian space, where
the dynamical properties can be extracted. It is not a general procedure which can be easily generalized to any amplitude, but it works fine for a
simple correlator which has well defined complex poles. It is a pragmatic approach which can be used to explore and clarify the physical meaning of the complex-conjugated
poles of the gluon propagator. As argued above, according to Refs.\cite{xigauge,excited,nielsen}, 
the whole principal part of the gluon propagator 
seems to be  gauge-parameter-independent
in a covariant gauge, pointing to a physical role played by poles and residues. 
The present analysis aims to clarify 
that role and might connect somehow the principal part to physical objects like condensates and observable 
two-particle gluon spectra.

\section{Wick rotation and standard analytic continuation}

Before going to the description of the anomalous case, 
it is useful to recover some known results in order to fix the notation.
Moreover, we can show how the usual relation between time orderings, in the Euclidean and Minkowski spaces,
emerges naturally as the unique choice which can be made.

As usual, the time-ordered gluon propagator is defined as
\begin{align}
i\,\Delta^{\mu\nu}(x-y)&=\langle 0\vert T\left\{A^\mu(x)A^\nu(y)\right\}\vert 0\rangle=\nn\\
&=\theta(x_0-y_0)\langle 0\vert A^\mu(x)A^\nu(y)\vert 0\rangle+\theta(y_0-x_0)\langle 0\vert A^\nu(y)A^\mu(x)\vert 0\rangle
\label{defDelta}
\end{align}
and its Fourier transform is given by
\BE
\Delta^{\mu\nu}(x)= \int\ppp\, \Delta^{\mu\nu}(p)\, e^{-ip\cdot x}
\label{AntiFourier}
\EE
\BE 
\Delta^{\mu\nu}(p)= \int {\rm d}^4x \, \Delta^{\mu\nu}(x)\, e^{ip\cdot x}. 
\label{Fourier}
\EE
The propagator can also be written in terms of two scalar functions, the transverse and longitudinal propagators,
\BE
\Delta^{\mu\nu}(p)=t^{\mu\nu}(p)\,\Delta^T(p^2)+\ell^{\mu\nu}(p)\,\Delta^L(p^2)
\EE
where $t^{\mu\nu}$, $\ell^{\mu\nu}$ are the transverse and longitudinal projectors, respectively.
In the Landau gauge the propagator is purely transversal and given by the function $\Delta(p^2)=\Delta^T(p^2)$.

Quite generally, the Euclidean function is usually obtained by Wick rotation, setting $p_0=ip_4$ so that $p^2=-p_E^2$. If the propagator 
has no poles in the first and third quadrant, then the Euclidean function is obtained by
\BE
\Delta_E(p_E^2)=\pm\Delta(-p_E^2),
\label{trivial}
\EE
where the sign is negative for a scalar field and positive for a vector field, because of the extra minus which arises when the
vectors $A^\mu$ are replaced by the Euclidean vector fields.

For instance, for a scalar field, replacing $A^\mu$ by $\Phi$ in Eq.(\ref{defDelta}), the free particle Feynman propagator is 
\BE
\Delta(p^2)=\frac{1}{p^2-m^2+i\epsilon}
\label{freescalar}
\EE
and the Euclidean function is
\BE
\Delta_E(p_E^2)=-\Delta(-p_E^2)=\frac{1}{p_E^2+m^2}.
\label{freeEscalar}
\EE
For the gluon, in the Landau gauge, the free particle Feynman propagator  is
\BE
\Delta^{\mu\nu}(p)=\frac{-t^{\mu\nu}(p)}{p^2+i\epsilon}
\label{free}
\EE
where
\BE
t^{\mu\nu}(p)=g^{\mu\nu}-\frac{p^\mu p^\nu}{p^2}.
\EE
Replacing $A^\mu$ and $p^\mu$ by the Euclidean vectors 
\BE
A^\mu A^\nu \left(g^{\mu\nu}-\frac{p^\mu p^\nu}{p^2}\right)= -A_E^\mu A_E^\nu \left(\delta^{\mu\nu}-\frac{p_E^\mu p_E^\nu}{p_E^2}\right)
\EE
(notice the presence of the extra minus sign)
so that replacing $t^{\mu\nu}$ by the Euclidean projector
\BE
t_E^{\mu\nu}=\delta^{\mu\nu}-\frac{p_E^\mu p_E^\nu}{p_E^2}
\EE
we obtain
\BE
\Delta_E^{\mu\nu}(p_E)=\frac{t_E^{\mu\nu}(p_E)}{p_E^2}
\label{freeE}
\EE
and for the transverse function
\BE
\Delta_E(p_E^2)=\Delta(-p_E^2)=\frac{1}{p_E^2}.
\label{freeET}
\EE

\vskip 10pt

For later use, we would like to discuss the same results in more detail, using the standard relations between correlators
in Minkowski and Euclidean space, which hold in perturbation theory, according to Eq.(\ref{circle}).
By the standard analytic continuation from real time to imaginary time, 
the Euclidean correlators follow by setting $x_0=t=-i\tau=-i\,x_4$ in Eq.(\ref{AntiFourier}) and Wick rotating the path of integration in the 
Fourier transform, using $p_0=i p_4$, yielding the Euclidean Fourier transform $\Delta_E$. More explicitly, the Fourier
transform provides an integral representation of the real time propagator which can be continued to imaginary time.
For a scalar field
\BE
\int_{-\infty}^{+\infty}\Delta(p)\, e^{-i p_0 x_0} \frac{{\rm d} p_0}{2\pi}=
\int_{-\infty}^{+\infty} \Delta(p)\, e^{-i p_4 x_4} \frac{i{\rm d}p_4}{2\pi}
\label{x4x0}
\EE
where the equality follows by rotating the integration path on the right-hand side,
while the extra $i$ factor ensures that the integrand function
$i\Delta(-p_E^2)$ is the Euclidean Fourier transform of the function 
\BE
\Delta({\bf x}, x_0=-ix_4)=-i\langle 0\vert T_\tau\left\{\Phi({\bf x}, -ix_4)\Phi(0)\right\}\vert 0\rangle
\label{Dtau}
\EE
so that $\Delta_E(p_E^2)=-\Delta(-p_E^2)$ is the Euclidean Fourier transform of the imaginary time correlator
\BE
\Delta_E(x)=\langle 0\vert T_\tau\left\{\Phi(x)\Phi(0)\right\}\vert 0\rangle
\label{Ecorr}
\EE
where the time ordering is on the imaginary time $\tau=x_4$. 

We observe that the integration path has been modified in
the Wick rotation, but the integral does not change if there are no poles in the first and third quadrant of the  complex $p_0$ plane.
Of course, Jordan's lemma ensures that the contour integrals vanish. Actually, for $x_0>0$ the exponential factor $\exp(-ip_0x_0)$ requires that the contour is closed in the lower halfplane (third and fourth quadrant), while for $x_4>0$ the exponential factor $\exp(-ip_4x_4)$ requires that the contour is closed in the right halfplane (for $\Im p_4<0$), including first and fourth quadrant of the complex $p_0$ plane, as shown on the left side of Fig.~1.
Thus, if we replace time ordering by imaginary-time ordering, and if there are no poles in the first and third quadrant, 
the integral does not change and is given by the residues of the fourth quadrant. The same argument works for $x_0<0$ ($x_4<0$).
For instance, the free particle propagator in Eq.(\ref{freescalar}) has  poles in the second and fourth quadrants of the complex $p_0$ plane, 
so that the Euclidean propagator is trivially obtained by Eq.(\ref{trivial}). The same result is found for the vector fields, 
with an extra minus sign from the Euclidean vectors $A_E^\mu A_E^\nu$, in agreement with Eq.(\ref{freeET}).

\begin{figure}[htb]
\includegraphics[width=0.9\textwidth]{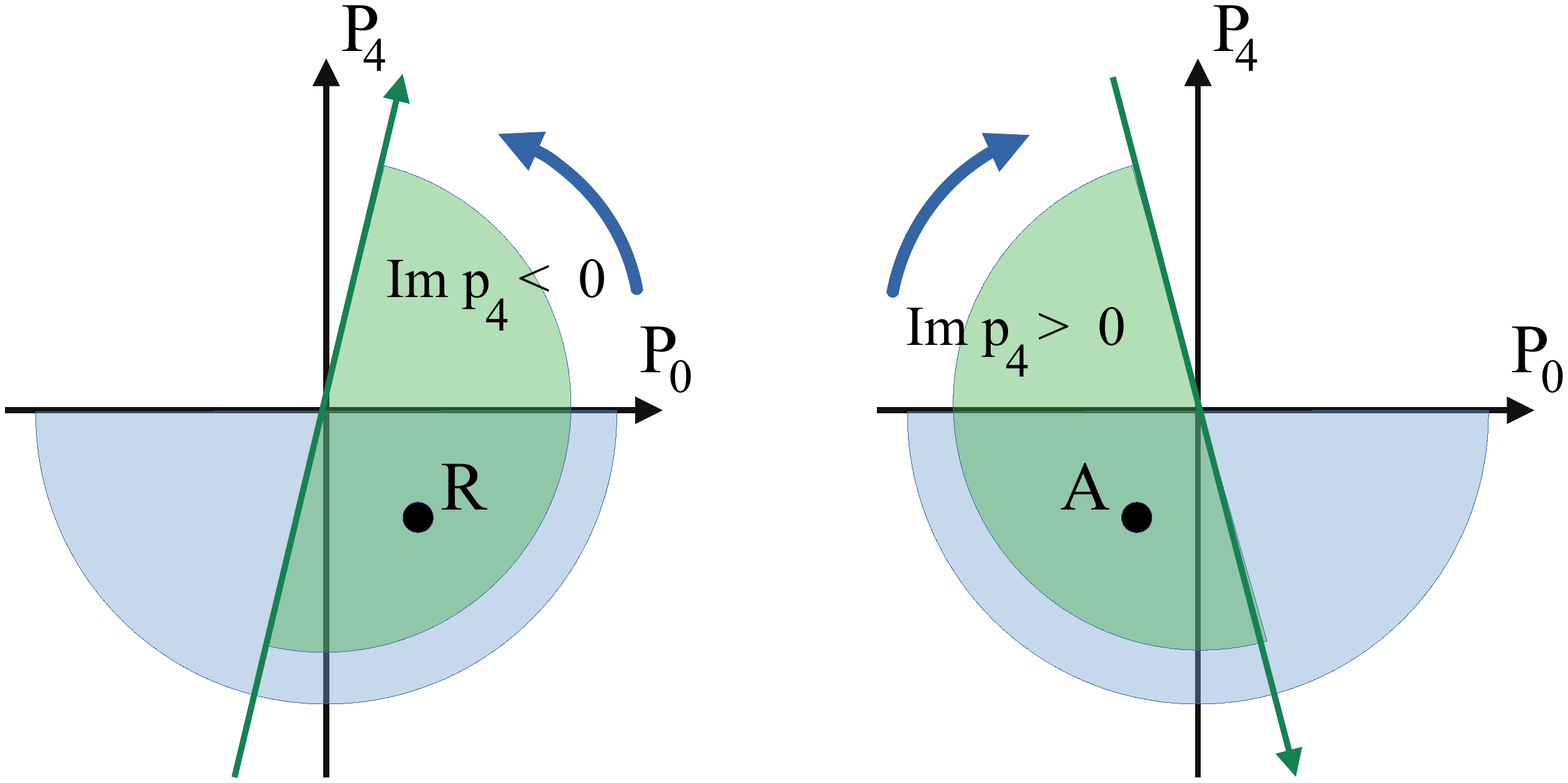}
\centering
\caption{On the left side, the usual anti-clockwise Wick rotation is shown, for a regular pole (R) which is in the fourth quadrant. 
On the right side, a clockwise Wick rotation is required for an anomalous pole (A) which is in the third quadrant. In both cases $x_0>0$. The shaded areas are the contours which must be chosen for $t>0$ in the Fourier transform, according to Jordan lemma.}
\label{Fig:F1}
\end{figure}

{\it We would like to stress that both, the rotation and the analytic continuation, must be taken together when the Fourier transform
is used as an integral representation of the real-time propagator, otherwise the integral would not be defined. Moreover, the imaginary time-order is enforced by the analytic continuation}.

In order to make the point clear, let us discuss the case of 
the free scalar field, Eq.(\ref{freescalar}), and write the integral representation
\BE
\Delta(t)=\int_{-\infty}^{+\infty}\frac{{\rm d} p_0}{2\pi}\, e^{-i p_0 t}\,\frac{1}{p_0^2-(\Omega-i\eta)^2}
\label{intrep1}
\EE
where $(\Omega-i\eta)^2={\bf p}^2+m^2-i\epsilon$ with $\Omega>0$, i.e. $\Omega=\sqrt{{\bf p}^2+m^2+\eta^2}$.
More generally, the imaginary part $\eta>0$ 
is not required to be small for the following discussion.
Let us assume that $t>0$ and evaluate the integral in the lower half-plane where it gives the finite result
\BE
\Delta(t)=-i\,\frac{e^{-i\Omega t}} {2(\Omega-i\eta)}\, e^{-\eta t}
\EE
which arises from the pole at $p_0=\Omega-i\eta$ in the fourth quadrant.
From the positive half-line $t>0$, the function $\Delta(t)$ can be continued to the lower half-plane according to
\BE
t=\vert t\vert e^{-i\theta}=a-ib
\EE
where $\theta$ can be increased continuously from zero to $\pi/2$, so that $a,b>0$. The analytic continuation reads
\BE
\Delta(a-ib)=-i\,\frac{e^{-i\Omega (a-ib)}} {2(\Omega-i\eta)}\, e^{-\eta (a-ib)}.
\label{contfunc}
\EE
It is immediately obvious that if the same continuation is taken in the integral representation of Eq.(\ref{intrep1}) 
the integral diverges: the exponential factor becomes $\exp(-i p_0 a-p_0 b)$ and diverges in the lower limit $p_0\to-\infty$. 
The analytic continuation of the integral can only be a representation of the continued function, Eq.(\ref{contfunc}),
{\it if}  the integration path is rotated 
anti-clockwise by the same angle
\BE
p_0= p_4 e^{i\theta}
\EE
where $p_4$ is a generic real variable which will become the fourth Euclidean component in the limit $\theta\to \pi/2$.
We obtain the modified integral representation
\begin{align}
\tilde\Delta(a-ib)&=\int_{-\infty}^{+\infty}\frac{{\rm d}  p_4\,e^{i\theta} }{2\pi}\, e^{-i p_4\vert t\vert}
\,\frac{1}{p_4^2\exp(2i\theta)-(\Omega-i\eta)^2}=\nn\\
&=e^{-i\theta}\int_{-\infty}^{+\infty}\frac{{\rm d}  p_4}{2\pi}\, e^{-i p_4\vert t\vert}
\, \left[\frac{1}{p_4-(\Omega-i\eta)\exp(-i\theta) }\right]\left[\frac{1}{p_4+(\Omega-i\eta)\exp(-i\theta) }\right].
\end{align}
The integral is finite and, since $\vert t\vert>0$, it can be evaluated in the lower half-plane of the complex variable $p_4$  
where the pole is at $p_4=(\Omega-i\eta)\exp(-i\theta)$. Since $\Omega>0$ and $\theta<\pi/2$ the pole is rotated by
an angle $-\theta$ but is still in the lower half-plane. By the same argument, the other pole remains in the upper half-plane.
Taking the contribution of the pole, the integral yields
\BE
\tilde\Delta(a-ib)=-ie^{-i\theta} \,\frac{e^{-i(\Omega-i\eta)\exp(-i\theta)\, \vert t\vert }} {2(\Omega-i\eta)\exp(-i\theta)}
=-i\frac{e^{-i(\Omega-i\eta)(a-ib)}} {2(\Omega-i\eta)}=\Delta(a-ib).
\label{intrep2}
\EE
Thus, the modified representation gives the correct analytic continuation of the propagator for any $\theta$ up to $\pi/2$.
The simultaneous rotations are necessary, at any value of $\theta$, in order to maintain the integral finite. Moreover,
denoting $t=-i\tau$, we observe that $\tau=b+ia$ and since $b>0$ then $\Rerm{\tau}>0$ in the present analytic continuation.
Should we have chosen $t<0$, the same argument would lead to $\Rerm{\tau}<0$. 

Of course, in the special case $\theta=\pi/2$ we recover the standard Wick rotation with $a\to 0$, $b=\tau>0$ and
$\Delta(-i\tau)$ which agrees with Eqs.(\ref{x4x0}),(\ref{Dtau})
\BE
\Delta(-i\tau)=\int_{-\infty}^{+\infty}\frac{{\rm d}  p_4 }{2\pi}\, e^{-i p_4\tau}\, \left[ i\Delta(-p_E^2)\right]
\label{intRE1}
\EE
where $p_E^2=p_4^2+{\bf p}^2$.
More generally, the same result, with the same imaginary-time ordering, is found whenever the poles of the propagator are 
all in the second and fourth quadrant, allowing the Wick rotation in the first and third quadrants, without encountering singularities. The argument is invalidated if there are {\it anomalous} singularities in the first and third quadrant.

\section{Anomalous poles and modified analytic continuation}

When complex conjugated poles are present, Eq.(\ref{trivial}) does not hold in general and the Wick rotation is not well defined.
The Euclidean (transverse) gluon propagator has been reported to have complex conjugated poles with a 
gauge-parameter independent principal part\cite{xigauge,nielsen}
\BE
\Delta_E(p_E^2)=\frac{R}{p_E^2+M^2}+\frac{R^\star}{p_E^2+{M^\star}^2}+\Delta\>_{\rm finite}
\label{glE}
\EE
with complex mass $M^2$ and residue $R$, and with real and imaginary parts which have the same sign: $\Im M^2>0$, $\Im R>0$, $\Rerm R>0$, 
$\Rerm M^2>0$\cite{xigauge}.

Even if the poles and residues seem to be gauge-parameter independent\cite{xigauge,nielsen}, 
it is not easy to understand their physical meaning. In fact, the analytic continuation to Minkowski space is not trivial because there are poles in all quadrants.
With the notation $M^2=m^2+i\gamma^2$ and denoting by $\omega^2={\bf p}\,^2+m^2$, neglecting the finite part, the Euclidean propagator reads
\BE
\Delta_E(p_E^2)=\frac{R}{p_4^2+\omega^2+i\gamma^2}+\frac{R^\star}{p_4^2+\omega^2-i\gamma^2}.
\label{gl4}
\EE
In the complex $p_0$ plane, denoting $p_0=ip_4$ as usual, the propagator reads
\BE
\Delta_E(p_E^2)=-\frac{R}{p_0^2-\omega^2-i\gamma^2}-\frac{R^\star}{p_0^2-\omega^2+i\gamma^2}
\label{gl0}
\EE
and there are poles at $p_0^2=\omega^2\pm i\gamma^2$. At variance with the free particle propagators of Eqs.(\ref{freescalar}),(\ref{free})
the poles are in all quadrants and the analytic continuation requires more care.

\subsection{Residue subtraction}

The problem of extracting the physical content in Minkowski space was addressed by Ref.\cite{detmold}.
In that work, the residue of the anomalous
pole was added to the calculation in Minkowski space if the trivially continued propagator was used. Here, we show that the procedure is equivalent to the definition of a new effective propagator in Minkowski space with a modified principal part.

Let us take the view that the propagator can be regarded as a distribution acting on physical well behaved functions $f$ 
which ensure convergence
\BE
I=\int_{-\infty}^{+\infty} \Delta_E(p_4) f(p_4) {\rm d}p_4=\int_{-\infty}^{+\infty} \Delta_E(x_4) f(-x_4) {\rm d}x_4
\EE
Here, we describe the scalar case and use a shorthand notation, omitting the integrals over ${\bf p}$ and over ${\bf x}$ in order to focus on the fourth component. 
An integral of that kind is encountered in the calculation of Feynman graphs and cross sections.
One usually assumes that all functions have no poles in the first and third quadrant, so that the integral does not
change when Wick rotating. The same physical observable can then be evaluated in Minkowski space by the change of variable
$p_4=-ip_0$ 
\BE
I=\int_{-\infty}^{+\infty} \Delta_E(p_4) f(p_4) {\rm d} p_4=\int_{-i\infty}^{+i\infty} \Delta_E(-ip_0) f(-ip_0) (-i{\rm d}p_0)
\label{variable}
\EE
followed by the usual clockwise rotation of the integration path, yielding
\BE
I=\int_{-\infty}^{+\infty} [-i \Delta_E(-ip_0)] f(-ip_0) {\rm d}p_0.
\label{clock}
\EE
Thus, the same physical content is obtained in Minkowski space if the function $i\Delta(p_0)$ is used in the graph,
with the Minkowskian propagator $\Delta(p_0)=-\Delta_E(p_4=-ip_0)$, recovering the same rule of Eq.(\ref{trivial}). 
An obvious extra minus sign occurs for the vector fields.

The same argument does not work if the propagator happens to have poles in the first and third quadrant. In that case,
the poles give contributions which must be added to the integral $I$ when the path of integration is modified\cite{detmold}. 

In more detail, let us enforce the conditions on the function $f$ and assume that the function has no poles in the
first and third quadrant, as it is often the case. 
We can split the principal part of the propagator in Eq.(\ref{gl0}), denoting by $\Delta_{A}$ the part which has {\it anomalous} 
poles at $p_0^2=\omega^2+i\gamma^2$ (first and third quadrant) and by $\Delta_{R}$ the {\it regular} part 
which has the usual poles at $p_0^2=\omega^2-i\gamma^2$ (second and fourth quadrant)

\begin{align}
\Delta_{A}(p_0)&=-\frac{R}{p_0^2-\omega^2-i\gamma^2}\nn\\
\Delta_{R}(p_0)&=-\frac{R^\star}{p_0^2-\omega^2+i\gamma^2}.
\label{split1}
\end{align}
The anomalous term $\Delta_{A}$ is the only part in the whole propagator which does not allow the usual Wick rotation. 
We observe that $\Delta_A(p_0)$ and $\Delta_R(p_0)$, despite the dependence on $p_0$ are here the {\it Euclidean} 
propagators, with the mere substitution $p_4=-ip_0$.

Let us see what happens to the anomalous case and take the Euclidean propagator $\Delta_E(p_4)$  equal to the anomalous
term of Eq.(\ref{split1})
\BE
\Delta_E(-ip_0)=\Delta_{A}(p_0)=-\frac{R}{p_0^2-\omega^2-i\gamma^2}
\label{DE13}
\EE
which has poles in the first and third quadrants, at $p_0=\pm z_0$, where $z_0=\sqrt{\omega^2+i\gamma^2}$ with $\Im z_0>0$, and residues
$\mp{R}/{(2z_0)}$, respectively.
By insertion in Eq.(\ref{variable}) we can write
\BE
I=\int_{-\infty}^{+\infty} \Delta_E(p_4) f(p_4) {\rm d} p_4
=\int_{-i\infty}^{+i\infty}\left[ -\frac{R}{p_0^2-\omega^2-i\gamma^2}\right]
f(-ip_0) (-i{\rm d}p_0).
\label{variable2}
\EE
In order to rotate (clockwise) the path of integration, we observe that the integrals along the imaginary and real axis differ by $2\pi i$ times
the residue of the integrand function, taken with opposite signs in the first and third quadrants, since the poles must be encircled in the opposite direction when deforming the path
\begin{align}
I&=\int_{-i\infty}^{+i\infty}\left[ -\frac{R}{p_0^2-\omega^2-i\gamma^2}\right]
f(-ip_0) (-i{\rm d}p_0)=\nn\\
&=\int_{-\infty}^{+\infty}\left[ -\frac{R}{p_0^2-\omega^2-i\gamma^2}\right]
f(-ip_0) (-i{\rm d}p_0)+2\pi \left[\frac{Rf(-iz_0)}{2z_0}+\frac{Rf(iz_0)}{2z_0}\right].
\label{clock2}
\end{align}
Quite generally, the function $f$ can be assumed to be even, since its odd part would give no contribution in the integral with an even
function $\Delta_E(p_4)$. Thus, the difference of the integrals is just $2\pi R f(iz_0)/z_0$. 
However, the integral on the real axis, on the right hand side of Eq.(\ref{clock2}), can be evaluated by closing the contour in the upper halfplane, yielding
\BE
\int_{-\infty}^{+\infty}\left[ -\frac{R}{p_0^2-\omega^2-i\gamma^2}\right]
f(-ip_0) (-i{\rm d}p_0)=2\pi \left[\frac{-R f(-iz_0)}{2z_0}\right]+C[f]
\EE
where $C[f]$ is the contribution coming from the poles of $f$ in the second quadrant. If we can neglect the extra term $C[f]$,
the integral on the right hand side of Eq.(\ref{clock2}) would be
just one half of the added term, but with the opposite sign. Thus the added term would reverse the sign of the integral
and we could write
\BE
I=\int_{-i\infty}^{+i\infty}\left[ -\frac{R}{p_0^2-\omega^2-i\gamma^2}\right]
f(-ip_0) (-i{\rm d}p_0)=\int_{-\infty}^{+\infty}\left[ \frac{R}{p_0^2-\omega^2-i\gamma^2}\right]
f(-ip_0) (-i{\rm d}p_0)
\label{clock3}
\EE
or, using Eq.(\ref{DE13})
\BE
I=\int_{-\infty}^{+\infty}\left[ i\Delta_E(-ip_0)\right] f(-ip_0) {\rm d}p_0
\EE
which has the opposite sign of Eq.(\ref{clock}).
We would conclude that the effective Minkowskian propagator has the opposite sign, $\Delta(p_0)=\Delta_E(p_4=-ip_0)$, compared with Eq.(\ref{trivial}), while the regular part of the propagator would maintain the same sign.
This result would be exact if the function $f$ had no poles at all, but then some convergence problems would arise. 
Actually, as we are going to show below, the argument can be made more rigorous if the function $f$ is just the Fourier exponential, with some limits on
convergence which can be used to establish the correct analytic continuation from real to imaginary time.
In the general case, in presence of other poles, the correct residues must be added by hand 
when going from the Euclidean to the Minkowski formalism, according to Eq.(\ref{clock2}),
as discussed in Ref.\cite{detmold}.

\subsection{Analytic continuation in time: clockwise and anti-clockwise rotation}

A reversing of sign in the anomalous part of the effective Minkowskian propagator can be recovered by going back to
real time and imposing that the analytic continuation to imaginary time satisfies 
a more general time ordering, as in Eq.(\ref{thetapm})  
\BE
\theta(t)\>\Longleftrightarrow \>\theta(\pm\tau)
\EE
where the sign might depend on the operators in the average. When $f(p_4)=\exp(-ip_4x_4)$ the integral in the previous section IIIA becomes the F.T. which defines the correlator in  imaginary-time. The function $f$ has no poles and the argument becomes exact {\it if} the convergence can be guaranteed by the Jordan lemma.

In the regular case, we can derive the Minkowskian propagator as the usual analytic continuation in time,
by just replacing $x_4=i x_0$ in the Euclidean correlator of Eq.(\ref{Ecorr}), inserting an extra minus
sign because of the vector fields and 
reversing the same steps of Eq.(\ref{x4x0}), yielding 
\begin{align}
\Delta^{\mu\nu}(x)&=i\Delta_E^{\mu\nu}({\bf x}, x_4=ix_0)=\nn\\
&=\int \frac{{\rm d}^3{\bf p}}{(2\pi)^3} e^{i{\bf p}\cdot{\bf x}}\int_{-\infty}^{+\infty} 
\Delta^{\mu\nu}_E({\bf p}, p_4=-ip_0)\, e^{-ip_0x_0} \frac{{\rm d}p_0}{2\pi}
\label{imaginarypath}
\end{align}
where the usual anti-clockwise Wick rotation, $p_0=ip_4$, must be taken together with the analytic continuation, 
in order to maintain a meaningful integral representation, as discussed
in detail through Eqs.(\ref{intrep1})-(\ref{intrep2}) and shown on the left side of Fig.~1.

Next, let us examine what happens to the anomalous term $\Delta_{A}$.
If there are poles in the first and third quadrant, the path of integration can only be clockwise rotated in 
order to reach the real axis without encountering singularities. The whole discussion of  section II still
holds, provided that the angle $\theta$ is replaced by $-\theta$, since $\eta$ must be replaced by $-\eta$
in the integral representation of the anomalous part, according to Eq.(\ref{intrep1}). 
Let us follow the same steps in detail: reversing the imaginary part of the poles, Eq.(\ref{intrep1}) reads
\BE
\Delta(t)=\int_{-\infty}^{+\infty}\frac{{\rm d} p_0}{2\pi}\, e^{-i p_0 t}\,\frac{1}{p_0^2-(\Omega+i\eta)^2}
\label{intrep3}
\EE
where $\eta>0$. We have anomalous poles in the first and third quadrant, at $p_0=\pm(\Omega+i\eta)$.
As before, let us assume that $t>0$ and evaluate the integral in the lower half-plane where it gives the finite result
\BE
\Delta(t)=i\,\frac{e^{i\Omega t}} {2(\Omega+i\eta)}\, e^{-\eta t}
\EE
which arises from the pole at $p_0=-(\Omega+i\eta)$ in the third quadrant, as shown on the right side of Fig.~1.

The function $\Delta(t)$ can be continued to the upper half-plane according to
\BE
t=\vert t\vert e^{i\theta}=a+ib
\EE
where $\theta$ can be increased continuously from zero to $\pi/2$, so that $a,b>0$. The analytic continuation reads
\BE
\Delta(a+ib)=i\,\frac{e^{i\Omega (a+ib)}} {2(\Omega+i\eta)}\, e^{-\eta (a+ib)}.
\label{contfunc2}
\EE
Again, the integral representation of Eq.(\ref{intrep3}) can only be meaningful if the integration path is rotated, together with
the analytic continuation, by the clock-wise rotation
\BE
p_0= p_4 e^{-i\theta},
\EE
as shown on the right side of Fig.~1,
yielding the modified integral representation
\begin{align}
\tilde\Delta(a+ib)&=\int_{-\infty}^{+\infty}\frac{{\rm d}  p_4\,e^{-i\theta} }{2\pi}\, e^{-i p_4\vert t\vert}
\,\frac{1}{p_4^2\exp(-2i\theta)-(\Omega+i\eta)^2}=\nn\\
&=e^{i\theta}\int_{-\infty}^{+\infty}\frac{{\rm d}  p_4}{2\pi}\, e^{-i p_4\vert t\vert}
\, \left[\frac{1}{p_4-(\Omega+i\eta)\exp(i\theta) }\right]\left[\frac{1}{p_4+(\Omega+i\eta)\exp(i\theta) }\right].
\label{intmod}
\end{align}
The integral is finite and, since $\vert t\vert>0$, it can be again evaluated in the lower half-plane of the complex variable $p_4$  where the pole is now at $p_4=-(\Omega+i\eta)\exp(i\theta)$. Since $\Omega>0$ and $\theta<\pi/2$ the pole is rotated by a positive angle $\theta$ but is still in the lower half-plane. By the same argument, the other pole remains in the upper half-plane.
Taking the contribution of the pole, the integral yields
\BE
\tilde\Delta(a+ib)=-ie^{i\theta} \,\frac{e^{i(\Omega+i\eta)\exp(i\theta)\, \vert t\vert }} {-2(\Omega+i\eta)\exp(i\theta)}
=i\frac{e^{i(\Omega+i\eta)(a+ib)}} {2(\Omega+i\eta)}=\Delta(a+ib).
\label{intrep4}
\EE
Thus, even in the anomalous case, the modified representation still gives the 
correct analytic continuation of the propagator for any $\theta$ up to $\pi/2$. However,
denoting $t=-i\tau$, this time we find $\tau=-b+ia$ and since $b>0$ then $\Rerm{\tau}<0$ in the 
analytic continuation, which is the opposite of the regular term.
Should we have chosen $t<0$, the same argument would lead to $\Rerm{\tau}>0$. 

Finally, in the special case $\theta=\pi/2$ we find $a\to 0$, $\tau=-b<0$ and Eq.(\ref{intmod}) reads
\BE
\Delta(-i\tau)=\int_{-\infty}^{+\infty}\frac{{\rm d}  p_4}{2\pi}\, e^{i p_4\tau}\, \left[- i\Delta(-p_E^2)\right]
\label{intRE2}
\EE
where $p_E^2=p_4^2+{\bf p}^2$ and the minus sign, inside the square brackets, arises from the opposite phase
$\exp(-i\theta)\to -i$.
In simple words, as shown on the right side of Fig.~1, 
the clock-wise rotation leads to an integral on the real axis going from $+\infty$ to $-\infty$,
so that a change of sign occurs. 

Reversing the steps, the anomalous term of the gluon propagator, $\Delta_A$ in Eq.(\ref{split1}), satisfies
\BE
\Delta^{\mu\nu}(x)=\int \frac{{\rm d}^3{\bf p}}{(2\pi)^3}e^{i{\bf p}\cdot{\bf x}}\int_{+\infty}^{-\infty} 
\Delta^{\mu\nu}_E({\bf p}, p_4=-ip_0)\, e^{-ip_0x_0} \frac{{\rm d}p_0}{2\pi}
\label{anomal}
\EE
which has the opposite sign of Eq.(\ref{imaginarypath}) because of the inversion of the extremes.
The sign of Eq.(\ref{trivial}) is reversed for the anomalous part 
\BE
\Delta^{\mu\nu}(p)=-\Delta^{\mu\nu}_E({\bf p}, p_4=-ip_0).
\EE
Moreover, the analytic continuation of the anomalous term
is only valid if $\tau<0$ when $t>0$, reversing the relation between time orderings.

We have seen that the convergence of the integrals enforces a strict relation between real and imaginary time ordering.
Actually, the integrals are not convergent if $x_4=x_0=0$ which
means that the propagator is not analytic at $t=\tau=0$. 
In the regular case, the analytic continuation works only if
a $\theta(t)$ corresponds to a $\theta(\tau)$, ensuring that the same pole is encircled in the integral representations,
in the Euclidean and Minkowski space, say Eq.(\ref{intRE1}) and Eq.(\ref{intrep1}), 
which give the same finite content because of
Jordan lemma. More generally,  the integrals are equivalent only if $\>\tau\, t>0$.
In the anomalous case, the time ordering is reversed: we still encircle the same pole in the integral representations,
say Eq.(\ref{intRE2}) and Eq.(\ref{intrep3}), which give the same finite content, provided that a $\theta(t)$ corresponds to a $\theta(-\tau)$,
or in other words, $\>\tau\, t<0$.

We conclude that, if the analytic continuation is enforced in time, then
the anomalous term $\Delta_{A}$ changes sign going form Euclidean to Minkowski space while
the normal term $\Delta_{R}$ does not.
With the notation of Eq.(\ref{split1}) the principal part of the {\it effective} Minkowskian propagator then reads
\BE
\Delta(p^2)=-\Delta_{A}(p_0)+\Delta_{R}(p_0)=\frac{R}{p_0^2-\omega^2-i\gamma^2}-\frac{R^\star}{p_0^2-\omega^2+i\gamma^2}
\EE
and using the definition of $\omega^2$
\BE
\Delta(p^2)=\frac{R}{p^2-M^2}-\frac{R^\star}{p^2-{M^\star}^2}.
\label{glM}
\EE
Then, when complex poles are present in the Euclidean propagator, 
the residue of the anomalous pole changes sign in Minkowski space, before taking the F.T.  which gives
the real time correlator, with important consequences on the the spectral properties.

For later reference, we observe that
neglecting the first (anomalous) term, and taking $\Im R=0$, $\Rerm R=Z$, $\gamma^2\to 0$, 
the propagator becomes the usual  massive gluon propagator in Minkowski space
\BE
\Delta(p^2)=\frac{Z}{-p^2+m^2-i\gamma^2}.
\label{Fey}
\EE

\subsection{Spectral properties}

Comparing Eq.(\ref{glE}) and Eq.(\ref{glM}), we observe that while the
Euclidean principal part is real on the real axis, the new Minkowskian principal part is a pure imaginary number. We are tempted
to see the pure imaginary principal part as a spectral density
\BE
\rho(p^2)=\frac{1}{i}\Delta(p^2)=\frac{1}{i}\left[\Delta_R(p^2)-\Delta_A(p^2)\right]
=\frac{1}{i}\left[\frac{R}{p^2-M^2}-\frac{R^\star}{p^2-{M^\star}^2}\right]
\EE
where we are using the same notation of Eq.(\ref{split1}):
\begin{align}
\Delta_{A}(p^2)&=-\frac{R}{p^2-M^2}\nn\\
\Delta_{R}(p^2)&=-\frac{R^\star}{p^2-{M^\star}^2}.
\label{split2}
\end{align}
Actually, the spectral density $\rho(p^2)$ has very interesting properties.
In fact, the regular part has no poles in the upper half-plane of the complex variable $p^2$,
then  satisfies the usual Kramers-Kronig dispersion relation
\BE
\Rerm \Delta_R(p^2)=\frac{1}{\pi}{\cal P}\int_{-\infty}^{+\infty}\frac{\Im\Delta_R(\mu^2)}{\mu^2-p^2} {\rm d}\mu^2
\label{disp1}
\EE
as can be easily confirmed by a direct calculation. Here, the imaginary part of $\Delta_R$  is just the spectral
weight $\rho$ since, on the real axis 
\begin{align}
\Delta_A(p^2)&=\left[\Delta_R(p^2)\right]^\star\nn\\
\Im\Delta_R(p^2)&=\frac{1}{2}\rho(p^2)\nn\\
\Delta_E(-p^2)&=\Delta_R(p^2)+\Delta_A(p^2)=2\Rerm \Delta_R(p^2)
\end{align}
where by $\Delta_E(-p^2)$ we mean the original Euclidean version of the principal part.
Thus, we can write
\BE
\Delta_E(-p^2)=\frac{1}{\pi}{\cal P}\int_{-\infty}^{+\infty}\frac{\rho(\mu^2)}{\mu^2-p^2} {\rm d}\mu^2
\label{disp2}
\EE
which holds on the real axis, strictly.

On the other hand, denoting by $\Delta^{f}_E(-p^2)$ the finite part $\Delta\>_{\rm finite}$
of the propagator in Eq.(\ref{glE}), 
it satisfies the usual K\"all\'en-Lehmann relation\cite{dispersion}:
\begin{align}
\Delta_E^{tot}(-p^2)=\Delta_A(-p^2)+\Delta_R(-p^2)+\Delta_E^{f}(-p^2)\nn\\
\Delta_E^f(-p^2)=\frac{1}{\pi}\int_0^\infty\frac{\Im \Delta_E^f(-\mu^2)}{\mu^2-p^2-i\epsilon} {\rm d}\mu^2
\label{KL}
\end{align}
where the spectral weight $\Im \Delta_E^f(-\mu^2)=0$ if $\mu^2<0$. Then, we can write in the Euclidean space ($p_E^2=-p^2>0$)
\BE
\Delta^{tot}_E(p_E^2)=\frac{1}{\pi}{\cal P}\int_{-\infty}^{+\infty}
\frac{\rho(\mu^2)+\Im \Delta_E^f(-\mu^2)}{\mu^2+p_E^2} {\rm d}\mu^2
\label{disp3}
\EE
where the spectral weight $\rho(\mu^2)$ adds the content of the principal part. The weight $\rho$ turns out to be the most
relevant contribution to the gluon propagator in actual calculations\cite{xigauge,dispersion}.

The integral representation of the principal part $\Delta_E(-p^2)$ in Eq.(\ref{disp2}) holds strictly on the real axis. For $p^2>0$
it gives the standard ``Minkowskian'' propagator which is obtained by a direct analytic continuation. However, this object is defined
on the real axis. We might define the same integral representation in the complex plane as
\BE
G(p^2)=\frac{1}{\pi}\int_{-\infty}^{+\infty}\frac{\rho(\mu^2)}{\mu^2-p^2} {\rm d}\mu^2
\label{G}
\EE
where $p^2$ is a generic complex variable. We can easily see that
\BE
G(p^2)= \begin{cases}
2\Delta_R(p^2) &\mbox{if } \Im p^2> 0 \\
\Delta_R(p^2)+\Delta_A(p^2) &\mbox{if } \Im p^2=0 \\
2\Delta_A(p^2) &\mbox{if } \Im p^2< 0.
\end{cases} 
\EE
The function $G(p^2)$ has no poles in the whole complex plane, but has a cut on the real axis where it jumps from $\Delta_R$ 
to $\Delta_A$. The difference on the cut gives the spectral function $\rho$. The function can be analytically continued across the cut on different Riemann sheets where the poles are found. We argue that, if a function like that could be
used in the Schwinger-Dyson equations, the proliferating of singularities and cuts which has been recently reported
in that formalism\cite{wink} could be somehow avoided or reduced.

Finally, the F.T. in Eqs. (\ref{anomal}) and (\ref{imaginarypath}) can be explicitly evaluated yielding the following
terms (omitting to indicate the three-vector integration on {\bf p})
\begin{align}
-i\Delta_A(t)&=-\frac{R}{2E}\left[\theta(t)\, e^{iE t}+\theta(-t)\, e^{-iE t} \right]\nn\\
-i\Delta_R(t)&=\frac{R^*}{2E^*}\left[\theta(t)\, e^{-iE^* t}+\theta(-t)\, e^{iE^* t} \right]
\label{ARt}
\end{align}
having introduced the complex energies $E^2={\bf p}^2+M^2$ with $\Im E >0$, $\Rerm E>0$. 
We observe that all the terms are well behaved 
and strongly damped in the limit $t\to\pm\infty$, as expected for a confined particle\cite{stingl,damp}.
In imaginary time, we find from the first integral in Eq.(\ref{imaginarypath}) 
\begin{align}
\Delta_A(\tau)&=\frac{R}{2E}\left[\theta(-\tau)\, e^{E \tau}+\theta(\tau)\, e^{-E \tau} \right]\nn\\
\Delta_R(t)&=\frac{R^*}{2E^*}\left[\theta(\tau)\, e^{-E^* \tau}+\theta(-\tau)\, e^{E^* \tau} \right].
\label{ARtau}
\end{align}
Again, all terms are well behaved in the limit $\tau\to\pm\infty$. We observe that the regular part $-i\Delta_R(t)$ is obtained by
the analytic continuation of the imaginary-time function $\Delta_R(\tau)$ with $t=-i\tau$ and with $\theta(\tau)$ replaced by
$\theta(t)$. The same analytic continuation, by $t=-i\tau$, also works for the anomalous function $\Delta_A(\tau)$ provided that
the sign of $\Delta_A(t)$ is changed and $\theta(\tau)$ is replaced by $\theta(-t)$. Thus the circle in Eq.(\ref{circle}) is
closed again having closed the chain by analytic continuations in time and energy. 

The Schwinger function behaves like $\exp(-\vert \tau\vert \Rerm E)$, where $\Rerm E>0$. In fact, we can write
\BE
\Delta_E(\tau)=\Delta_R(\tau)+\Delta_A(\tau)=
\left\{\theta(\tau)\left[\frac{R}{2E}\, e^{-E\tau}+\frac{R^*}{2 E^*}\, e^{-E^* \tau}\right]
+ \left(\tau\>\longleftrightarrow\> -\tau \right)\right\}
\EE
and taking $E=M$ in the limit ${\bf p}=0$, we obtain for $\tau>0$
\BE
\left[\Delta_E(\tau)\right]_{{\bf p}=0}\sim  \exp(-\tau\Rerm M)\,\cos(\phi-\tau\Im M)
\label{schwing}
\EE
where $\Rerm M>0$ and the phase $\phi$ is the difference between the arguments of $R$ and $M$,
i.e. $\phi=\arctan{(\Im R/\Rerm R)}-\arctan{(\Im M/\Rerm M)}\approx 0.69$ according to the data of Ref.\cite{xigauge}.
The Schwinger function becomes negative at 
$\tau=(\pi/2+\phi)/\Im M\approx 2.26/(0.375\, {\rm GeV})=6.0\,{\rm GeV}^{-1}\approx 1.2$~fm, where again, the data
of Ref.\cite{xigauge} have been used. This length-scale is consistent with the expected confinement radius of a gluon.
We observe that this prediction is gauge-parameter-independent,
as previously conjectured in Ref.\cite{alkofer04} if, and only if,
the phase of the residues and the poles are also invariant.

The real-time propagator is
\BE
-i\Delta(t)=-i\left[\Delta_R(t)-\Delta_A(t)\right]=
-i\left\{\theta(t)\left[\frac{R}{2 E}\, e^{i E t}+\frac{R^*}{2 E^*}\, e^{-i E^* t}\right]
+ \left(t\>\longleftrightarrow\> -t \right)\right\}.
\label{tprop}
\EE
and behaves like $\exp(-\vert t\vert \Im E)$, where $\Im E>0$. The two functions, $\Delta(t)$ and $\Delta_E(\tau)$,  
are not related by a trivial analytic continuation. It is important to observe that the reversed sign in the anomalous part
gives a natural aspect to the real-time propagator which can be written in terms of intermediate-state amplitudes, as discussed in the next section.

\section{Anomalous spectral representation}
In presence of complex conjugated poles, the K\"all\'en-Lehmann representation does not hold. The principal part must be added to the
usual dispersion relation\cite{dispersion,kondo18}. The added part has been seen in Eq.(\ref{disp3}) as deriving from the spectral weight $\rho(p^2)$ which is the Minkowskian principal part of the propagator, according to our procedure for going from Euclidean to
Minkowski space.

The gauge-parameter independence of the principal part\cite{xigauge,nielsen} and the relevance of the weight $\rho$ (which is the larger term in the total spectral function\cite{dispersion}) suggest that the residues and the poles might be related to a
phenomenologically relevant sector of the single-particle spectrum. For a confined gluon, the principal part could have the same role which is usually played by a real pole for an observable particle. 

If the principal part arises from an anomalous sector of the spectrum, we can extract some properties of that sector 
by a more detailed study of the Minkowskian principal part.  Moreover, Eq.(\ref{tprop}) suggests that the effective
propagator, with its anomalous sign, could be naturally related to a set of intermediate states, even if,
at variance with the usual single-particle pole of Eq.(\ref{Fey}), the principal part is a
pure imaginary number on the real axis. We are assuming that, according to several studies\cite{xigauge,GZ}, 
the constraints $\Im M^2>0$, $\Im R>0$, $\Rerm R>0$,  $\Rerm M^2>0$, are satisfied.

On general grounds, the propagator is defined as in Eq.(\ref{defDelta}) which can be written as
\begin{align}
i\,\Delta^{\mu\nu}({\bf x},t)&=\theta(t)\langle 0\vert A^\mu(0)e^{i{\bf P}\cdot {\bf x}}\,U(t)\, A^\nu(0)\vert 0\rangle
+\theta(-t)\langle 0\vert A^\nu(0)\,e^{-i{\bf P}\cdot {\bf x}} U(-t)\, A^\mu(0)\vert 0\rangle
\label{DeltaU}
\end{align}
where ${\bf P}$ is the momentum operator and $U(t)$ is the time-evolution operator. Without any special hypothesis on time evolution,
we can always write the elements of the group in terms of a generator $H$ which we {\it call}  Hamiltonian.
\BE
U(t)=e^{-iHt}.
\EE
If the space of states is a pseudo-Euclidean space, with negative-norm states, then the definition of Hermitian conjugation might require
some extra care. Here, we take the usual definition and say that $B^\dagger$ is the adjoint of the operator $B$ if
\BE
\langle \Psi\vert B^\dagger \vert \Phi\rangle=\langle  \Phi\vert B\vert\Psi\rangle^*
\EE
for any pair of states $\vert \Phi\rangle$, $\vert \Psi\rangle$, irrespective of their norm. In order to find a real 
expectation value of the Hamiltonian, one usually requires it to be a self-adjoint (Hermitian) operator, so that the set of
operators $U(t)$ give a unitary representation of the time-evolution group.
However, in presence of negative-norm states, the eiegnvalues of the Hermitian operator $H$ might not be real, and non-unitary
representations of $U(t)$ might exist. A simple example is provided in the next section. Denoting by $\{ {\cal P}_n\}$ a set of projectors
on eigenstates of $H$ with eigenvalues $E_n$, without any special assumption on the nature of the numbers $E_n$, we can write
\BE
U(t)\,{\cal P}_n={\cal P}_n\, U(t)={\cal P}_n\, e^{-i E_n t}
\EE
and the propagator reads
\BE
i\,\Delta^{\mu\nu}({\bf x},t)=\theta(t)\sum_n \rho^{\mu\nu}_n\> e^{i{\bf p}_n\cdot {\bf x}}\> e^{-iE_nt}
+\theta(-t)\sum_n \rho^{\nu\mu}_n \>e^{-i{\bf p}_n\cdot {\bf x}}\> e^{iE_nt}
\label{deltarho}
\EE
where
\BE
\rho^{\mu\nu}_n=\langle 0\vert A^\mu(0)\,{\cal P}_n\, A^\nu(0)\vert 0\rangle.
\EE
We observe that since, in general, ${\cal P}_n^\dagger\not={\cal P}_n$, the spectral weight $\rho^{\mu\nu}_n$ might even be a complex number.

Since we limit the study to the principal part, then the sum over the intermediate states can be regarded as a partial sum
over a limited subset which is not required to be complete. We argue that the principal part must arise from an anomalous
subset and yet share the same structure of Eq.(\ref{deltarho}). In fact, any {\it regular} subset would give a term which satisfies the 
standard K\"all\'en-Lehmann representation. As discussed in Ref.\cite{dispersion}, the principal part must be added by hand
to the usual dispersion relations, so that its contribution in Eq.(\ref{deltarho}) must arise from a special set of states which are not
present in the standard spectral representation. Thus, we only need to consider that special subset for the study of the principal part,
which we denote by $\Delta^{\mu\nu}$  from now on.

The Fourier transform gives
\BE
i\Delta^{\mu\nu}({\bf p},p_0)=\sum_n (2\pi)^3\delta^3({\bf p}-{\bf p}_n)\>\rho^{\mu\nu}_n \int_0^\infty e^{i(p_0-E_n)t} {\rm d}t
+\sum_n (2\pi)^3\delta^3({\bf p}+{\bf p}_n)\>\rho^{\nu\mu}_n\int_{-\infty}^{\>0} e^{i(p_0+E_n)t} {\rm d}t
\EE
and introducing a transverse projection 
\BE
\rho_n(p)=\rho_n(-p)=\frac{1}{d-1}t_{\mu\nu}(p)\rho^{\mu\nu}_n 
\EE
we can write the principal part of the transverse propagator as
\BE
i\Delta({\bf p},p_0)=\sum_n (2\pi)^3\delta^3({\bf p}-{\bf p}_n)\>\rho_n(p) \int_0^\infty e^{ip_0t}e^{-iE_n t} {\rm d}t
+\sum_n (2\pi)^3\delta^3({\bf p}+{\bf p}_n)\>\rho_n(p)\int_{0}^{\infty} e^{-ip_0t}e^{-iE_nt} {\rm d}t.
\EE
Because of parity, we can assume the existence of degenerate pairs of states with $E_n=E_{n^\prime}$ and
$\rho_n({\bf p},p_0)=\rho_{n^\prime}(-{\bf p},p_0)$
and write
\BE
i\Delta({\bf p},p_0)=\sum_n (2\pi)^3\delta^3({\bf p}-{\bf p}_n)\>\rho_n(p) \int_0^\infty 
\left[e^{ip_0t}+e^{-ip_0t}\right]\>e^{-iE_n t} {\rm d}t
\label{Dp}
\EE
where the symmetry $\Delta({\bf p},p_0)=\Delta({\bf p},-p_0)$, which follows from the Lorentz invariance 
of the principal part $\Delta(p)$  in Eq.(\ref{glM}), is made manifest yielding an even function
of $p_0$ and ${\bf p}$.

Complex conjugated poles can only arise if the energies $E_n$ are complex numbers. Thus, the first problem we must
face is the origin of complex eigenvalues for the Hermitian Hamiltonian $H$. Moreover, denoting by $\chi_n(x)$ the
{\it wave functions} in Eq.(\ref{deltarho})
\BE
\chi_n(x)=e^{i{\bf p}_n\cdot{\bf x}} e^{-iE_n t}
\EE
we observe that, because of Lorentz invariance, the quantity $\partial_\mu \partial^\mu\chi_n(x)$ must be a scalar and
then, the wave functions satisfy the Klein-Gordon (KG) equation on {\it complex mass shell}
\begin{align}
\partial_\mu\partial^\mu \> \chi_n(x)&=({\bf p_n}^2-E_n^2)\>\chi_n(x)=-M_n^{2}\>\chi_n(x)\nn\\
\partial_\mu\partial^\mu \> \chi_n^*(x)&=({\bf p_n}^2-E_n^{*\,2})\>\chi^*_n(x)=-M_n^{*2}\>\chi^*_n(x)
\end{align}
where the complex conjugated masses $M_n^2$, $M_n^{*\,2}$ must be Lorentz scalars. Then, the complex eigenvalues can
take the four different values
\BE
E_n=\pm \sqrt{{\bf p_n}^2+M_n^2},\qquad E_n^*=\pm \sqrt{{\bf p_n}^2+M_n^{*\,2}}.
\label{shell}
\EE
Denoting by 
\BE
\omega_n+i\gamma_n= \sqrt{{\bf p_n}^2+M_n^2}
\EE
with $\omega_n>0$, $\gamma_n>0$ and $\Im M_n^2>0$, the four energies can be written as the two pairs
\begin{align}
E_n&=\pm \omega_n+i\gamma_n\nn\\
E_n&=\pm \omega_n-i\gamma_n
\label{pairs}
\end{align}
but only the second pair can be accepted in Eq.(\ref{Dp}), since the first pair would give a divergent integral over time.
Thus, a second problem to be faced is the origin of intermediate states with energies $E_n=-\omega_n-i\gamma_n$ and 
$-E_n^*=\omega_n-i\gamma_n$, with negative imaginary parts. Actually, these energies are precisely the pair of frequencies
$-E$, $E^*$ that we have found in Eq.(\ref{tprop}) for $t>0$.
In the next section, a quite speculative toy model is discussed, where such eigenvalues arise 
from a mixing of positive- and negative-norm
states.

Here, in order to make sense of Eq.(\ref{Dp}), we just assume that a subset of such intermediate states does exist, 
with energies $E_n$, $-E_n^*$ sharing the same (real) spectral weight $\rho_n$ and the same negative imaginary part $\Im E_n=-\gamma_n$.
Under such assumptions the explicit integral in Eq.(\ref{Dp}) yields
\BE
\Delta(p)=(2\pi)^3\sum_n\rho_n(p)\,\delta^3({\bf p}-{\bf p_n})\>\left\{
\frac{1}{p_0-E_n}-\frac{1}{p_0+E_n}
+\frac{1}{p_0+E^*_n}-\frac{1}{p_0-E^*_n}\right\}.
\EE
Adding the terms and using the complex mass shell Eq.(\ref{shell})
\BE
\Delta(p)=\sum_n \frac{(2\pi)^3 \left[2\rho_n(p)\,E_n\right]\delta^3({\bf p}-{\bf p_n})}
{p^2-M_n^{2}}
-\sum_n \frac{(2\pi)^3 \left[2\rho_n(p)\,E_n^*\right] \delta^3({\bf p}-{\bf p_n})}
{p^2-M_n^{*\,2}}.
\label{Minkstruct}
\EE
The propagator is a pure imaginary number on the real axis and has the same structure of
the principal part discussed in the previous Section.

Restricting the sum to a single set of states with $M_n^2=M^2$, by Lorentz invariance we can write the result
as
\BE
\Delta(p)=\frac{R(p^2)}
{p^2-M^{2}}
-\frac{R^*(p^2)}
{p^2-M^{*\,2}}
\label{Minkp}
\EE
where the phase of the complex function $R(p^2)$ arises from 
the sum over the complex spectrum in Eq.(\ref{Minkstruct})
and must be gauge invariant
if the spectrum is assumed to be invariant.
If the function $R(p^2)$ is regular at the complex point $p^2=M^2$, then $R=R(M^2)$ is the residue and the principal
part reads
\BE
\Delta(p)=
\frac{R}{p^2-M^2}-\frac{R^*}{p^2-M^{2\,*}}
\EE
which is precisely the Minkowskian principal part which we found in Eq.(\ref{glM}) by the anomalous
analytic continuation.

From our knowledge of the gluon propagator\cite{xigauge,duarte}, the real and imaginary parts of $R$ are positive
($\Rerm R>0$, $\Im R>0$) and we assumed that $E_n=-\omega_n-i\gamma_n$, so that $\Rerm E_n<0$ and $\Im E_n<0$. Thus,
the spectral coefficient $\rho_n$ is expected to be negative. 

We observe that a negative $\rho_n<0$ is usually found for physical transversal states with a positive norm. In fact,
denoting by $b^\mu$ the vector 
\BE
b^\mu=t^{\mu\nu}\,  \langle 0\vert A_\nu(0)\vert n\rangle
\EE
we can write
\BE
\rho_n=\frac{1}{d-1}\>b^\mu b_\mu^*.
\EE
Because of its transversality, $b^\mu$ is a space-like vector if $p^2>0$. For instance, in a frame where $p^\mu=(p_0, 0, 0, p^3)$
with $(p_0)^2>(p_3)^2$, the vanishing of $p^\mu b_\mu$ gives $b_0=b_3 (p^3/p_0)$  and $\vert b_0\vert^2<\vert b_3\vert^2$, so that
$b^\mu b_\mu^*<0$.

We conclude that the Minkowskian principal part of the gluon propagator is compatible with the existence of a sub-set of anomalous
intermediate states with a positive norm, but with complex eigenvalues $\pm\omega-i\gamma$, 
sharing the same negative imaginary part.
Having a positive norm, the intermediate states might have a physical relevance, as prompted 
by the gauge-parameter independence of the principal part. Of course, we are far from having reached a 
valid microscopic proof of existence for such states. A speculative toy model which might predict the existence
of such scenario is discussed in the next section.

\section{A basic Hermitian model with complex eigenvalues}

Negative-norm states appear in the Gupta-Bleuler approach to the quantization of the electromagnetic field. While the unphysical intermediate states should be canceled by the ghost fields in the perturbative approach to QCD, the cancellation might not be as effective in the non-perturbative limit, where the gluon acquires a dynamical mass while the ghost seems to be massless.

In this section, by a very basic model, we show that if states with positive and negative norm are somehow coupled, in a pseudo-Euclidean space, an Hermitian Hamiltonian can have complex eigenvalues without affecting the unitarity of the time evolution.
While a more detailed discussion can be found in Ref.\cite{kondo21},
here we explore the consequences of a coupling between a physical state and a state with a negative norm. The discussion is quite general and could describe the coupling between different polarizations of a gluon and ghosts or even different fields. We assume that a physical,
positive-norm state and an unphysical degree of freedom, with a negative norm,  can mix because of the 
interactions in the non-perturbative vacuum. 

As a simple toy model, let us study the space spanned by just two one-particle states,
sharing the same momentum ${\bf k}$ and other quantum numbers.
We assume that the other polarizations can be regarded as decoupled. Omitting the label ${\bf k}$ and all other quantum numbers, by appropriate normalization, we denote the states by
\begin{align}
\vert \Phi_0\rangle&=a_0^\dagger \vert 0\rangle  \nn\\
\vert \Phi_1\rangle&=a_1^\dagger \vert 0\rangle  
\end{align}
where $\vert 0\rangle$ is the vacuum and we assume the following commutation relations
\begin{align}
[a_0, a_0^\dagger]&=-1,\qquad  [a_1, a_1^\dagger]=+1        \nn\\
[a_1, a_0^\dagger]&=[a_0, a_1^\dagger]=0  
\end{align}
while the normal-ordered Hamiltonian reads (without interactions)
\BE
H=\omega\left(a_1^\dagger a_1-a_0^\dagger a_0\right)
\label{hamilt}
\EE
which is the standard result for the free-particle Hamiltonian of a gauge field when projected in the subspace. 
Despite the minus sign, the states $\vert \Phi_0\rangle$, $\vert \Phi_1\rangle$  are eigenstates of $H$ with equal and positive
eigenvalue $\omega$. The eigenvalues might be even different (e.g. different masses) without affecting the main results. 
While the diagonal matrix element of $H$ is negative
\BE
\langle \Phi_0\vert H \vert \Phi_0\rangle=-\omega\langle \Phi_0\vert a_0^\dagger a_0\vert \Phi_0\rangle=
-\omega\langle 0\vert a_0 a_0^\dagger a_0 a_0^\dagger\vert 0\rangle=\omega\langle 0\vert a_0 a_0^\dagger\vert 0\rangle=-\omega
\EE
the average of $H$ is positive because of the negative norm of $\vert \Phi_0\rangle$
\begin{align}
\langle \Phi_0\vert\Phi_0\rangle&=\langle 0\vert a_0 a_0^\dagger\vert 0\rangle=-1\nn\\
\langle \Phi_1\vert\Phi_1\rangle&=\langle 0\vert a_1 a_1^\dagger\vert 0\rangle=+1
\end{align}
and then
\BE
\frac{\langle \Phi_0\vert H \vert \Phi_0\rangle}{\langle \Phi_0\vert \Phi_0\rangle}=
\frac{\langle \Phi_1\vert H \vert \Phi_1\rangle}{\langle \Phi_1\vert \Phi_1\rangle}=\omega.
\EE
We can also work out the time evolution of the operators by Heisenbeg equation
\begin{align}
i\frac{\partial}{\partial t} a_0&=[a_0, H]=-\omega[a_0,a_0^\dagger]a_0=\omega\, a_0 \quad \to\quad a_0(t)=a_0(0)e^{-i\omega t}\nn\\
i\frac{\partial}{\partial t} a_1&=[a_1, H]=\omega[a_1,a_1^\dagger]a_1=\omega\, a_1 \quad \to\quad a_1(t)=a_1(0)e^{-i\omega t}.
\end{align}

\vskip 10pt

We can introduce a matrix formalism for the two-dimensional pseudo-euclidean sub-space spanned by the set {$\vert\Phi_i\rangle$}
denoting by the vector $X^\mu$ the generic state
\BE
\vert X\rangle= X^1 \vert\Phi_1\rangle+X^0\vert \Phi_0\rangle=
\begin{pmatrix} 
 X^1 \\
 X^0 
\end{pmatrix}
\EE
and by $g_{\mu\nu}$ the two-dimensional pseudo-euclidean metric $g_{10}=g_{01}=0$, $g_{00}=-1$, $g_{11}=1$, i.e.
\BE
g_{\mu\nu}=\langle \Phi_\mu\vert\Phi_\nu\rangle=
\begin{pmatrix} 
1 & 0 \\
0 & -1 
\end{pmatrix}.
\EE

The scalar product between two states reads
\BE
\langle X\vert Y\rangle= X^{\mu*}g_{\mu\nu} Y^\nu= X_\nu^* Y^\nu
\EE
where the matix $g_{\mu\nu}$ is used for raising and lowering indices. In this formalism the Hamiltonian reads
\BE
H_{\mu\nu}=
\begin{pmatrix} 
\omega & 0 \\
0 & -\omega 
\end{pmatrix}
\EE
Of course, the matrix must be Hermitian in order to ensure the reality of the expectation value
\BE
\langle X\vert H\vert X\rangle= X^{\mu*}H_{\mu\nu} X^{\nu}=\langle X\vert H\vert X\rangle^*.
\EE
It is important to observe that the eigenvalues are not the diagonal elements of $H_{\mu\nu}$ but can be found on
the diagonal of $H^\mu_{\>\>\nu}$. In fact, the eigenvalue problem reads
\BE
H^\mu_{\>\>\nu} X^\nu=\lambda X^\mu\quad\to \quad (H^\mu_{\>\>\nu}-\lambda\delta^\mu_\nu)X^\nu=0
\EE
or
\BE
H_{\mu\nu} X^\nu=\lambda X_\mu\quad\to \quad (H_{\mu\nu}-\lambda g_{\mu\nu})X^\nu=0.
\EE
Actually, raising an index by $g^{\mu\nu}$ we find
\BE
H^\mu_{\>\>\nu}=
\begin{pmatrix} 
\omega & 0 \\
0 & \omega 
\end{pmatrix}
\EE
and $\omega$ is the correct eigenvalue for both states. 
\vskip 10pt

\subsection{Complex eigenvalues from interactions}

Because of the interactions, in the non-perturbative regime, the Hamiltonian might acquire an off-diagonal term that couples the
two states. We mimic such interaction by adding an Hermitian off-diagional term to
the free-particle Hamiltonian
\BE
H_{int}=\gamma \left[ e^{-i\theta} a_0^\dagger a_1+e^{i\theta} a_1^\dagger a_0\right]
\label{Hint}
\EE
where $\gamma>0$ depends on the coupling strength and $\gamma\to 0$ in the perturbative asymptotic limit.
We observe that $\gamma=0$ in the Gupta-Bleuler approach to QED, since there is no interaction between photons and
the free-particle states are decoupled asymptotic states. In the Lagrangian formalism, the ghosts are decoupled in QED
and there is  no mixing with the other degrees of freedom.  On the other hand, QCD is an intrinsically coupled theory with no
free-particle asymptotic states, then the eigenstates of the fully interacting theory are expected to be given by a superposition
of the free-particle states, or in other words, an off-diagonal term $\gamma\not=0$ must be present in the Hamiltonian.
In the Lagrangian formalism, there are ghosts which do not decouple in a covariant gauge. Their interaction with the gluon
also contributes to determine a set of unknown non-perturbative states which are hardly written as decoupled free-particle
states.    Then, some off-diagonal term must be added to the Hamiltonian of the toy-model in order to mimic the behavior
of the interacting QCD, at variance with QED.
Here, we do not investigate further the origin of the mixing, but just assume that
the off digonal term (\ref{Hint}) is present in the model Hamiltonian.
The total Hamiltonian matrix then reads
\BE
H_{\mu\nu}=
\begin{pmatrix} 
\omega & \gamma e^{i\theta} \\
\gamma e^{-i\theta} & -\omega 
\end{pmatrix}.
\EE
While this  Hamiltonian is Hermitian, the matrix $H^\mu_{\>\>\nu}$ is not
\BE
H^\mu_{\>\>\nu}=
\begin{pmatrix} 
\omega & \gamma e^{i\theta}\\
-\gamma e^{-i\theta} & \omega 
\end{pmatrix}.
\EE
The eigenvalues are
\BE
\lambda_{\pm}=\omega\pm i\gamma
\EE
and the eigenvectors can be written as
\BE
{\epsilon^{\pm}}^\mu=\frac{1}{\sqrt{2}}
\begin{pmatrix} 
e^{i\theta}\\
\pm i 
\end{pmatrix}.
\EE
It is remarkable that complex conjugated eigenvalues emerge from the mixing with negative-norm states even
if the Hamiltonian is Hermitian\cite{kondo21}.
The eigenvectors satisfy the properties
\begin{align}
H^\mu_{\>\>\nu}\> {\epsilon^{\pm}}^\nu&=\lambda_\pm\>{\epsilon^{\pm}}^\mu \nn\\
{\epsilon^{\pm}}^*_\mu\> H^\mu_{\>\>\nu} &=\lambda_\mp\> {\epsilon^{\pm}}^*_\nu 
\end{align}
i.e. they are eigenvectors on the left with the complex conjugated eigenvalue $\lambda_\mp=\lambda_\pm^*$.
Their norm is zero
\BE
{\epsilon^{\pm}}^*_\mu\>{\epsilon^{\pm}}^\mu=0
\EE
but their mixed product is
\BE
{\epsilon^{\pm}}^*_\mu\>{\epsilon^{\mp}}^\mu=1
\EE
so that two projectors can be built and the identity can be written as
\BE
\delta^\mu_\nu={\epsilon^{-}}^\mu\>{\epsilon^{+}}^*_\nu+{\epsilon^{+}}^\mu\>{\epsilon^{-}}^*_\nu.
\label{proj1}
\EE
Finally, the Hamiltonian has the spectral representation
\BE
H^\mu_{\>\>\nu}=\lambda_-\>{\epsilon^{-}}^\mu\>{\epsilon^{+}}^*_\nu\>+\>\lambda_+\>{\epsilon^{+}}^\mu\>{\epsilon^{-}}^*_\nu
\EE
and we can check that $H_{\mu\nu}$ is Hermitian
\BE
H_{\mu\nu}=\lambda_-\>{\epsilon^{-}}_\mu\>{\epsilon^{+}}^*_\nu\>+\>\lambda_+\>{\epsilon^{+}}_\mu\>{\epsilon^{-}}^*_\nu.
\EE

Denoting by $\vert \pm\rangle$ the eigenvectors ${\epsilon^{\pm}}^\mu$ and by $\langle \pm \vert$ their conjugate ${\epsilon^{\pm}}^*_\mu$,
all previous results can be written as
\begin{align}
H\vert \pm\rangle&=\lambda_\pm\>\vert \pm\rangle\nn\\
\langle \pm\vert H&=\lambda_\mp\>\langle\pm\vert
\end{align}
\begin{align}
\langle +\vert+\rangle&=\langle -\vert-\rangle=0\nn\\
\langle +\vert-\rangle&=\langle -\vert+\rangle=1
\end{align}
\begin{align}
I&=\vert-\rangle\langle+\vert\>+\>\vert+\rangle\langle -\vert\nn\\
H&=\vert-\rangle \lambda_- \langle +\vert\>+\>\vert+\rangle \lambda_+ \langle -\vert.
\label{proj2}
\end{align}

It is useful to define the creation operators $a_\pm^\dagger$ as

\BE
a_\pm^\dagger =
\frac{1}{\sqrt{2}}\left[e^{i\theta}a_1^\dagger\pm i a_0^\dagger\right]
\label{apm}
\EE
so that
\BE
a_\pm^\dagger \vert 0\rangle=
\frac{1}{\sqrt{2}}\left[e^{i\theta}a_1^\dagger\pm i a_0^\dagger\right]\vert 0\rangle=\vert \pm\rangle.
\EE
It is easy to see that they satisfy the commutation relations
\begin{align}
[a_+, a_+^\dagger]&=[a_-, a_-^\dagger]=0\nn\\
[a_-, a_+^\dagger]&=[a_+, a_-^\dagger]=1
\end{align}
and the Hamiltonian reads
\BE
H=\lambda_-\> a_-^\dagger a_+\>+\>\lambda_+\> a_+^\dagger a_-
\EE
and is obviously Hermitian, while retaining complex conjugated eigenvalues. The result can be recovered by inverting the definition
of $a_\pm^\dagger$ in Eq.(\ref{apm}) and inserting in Eq.(\ref{hamilt}).

For further reference, we can check the time dependence of the operators, as we did for the diagonal case:
\begin{align}
i\frac{\partial}{\partial t} a_+&=[a_+, H]=\lambda_-[a_+,a_-^\dagger]a_+=\lambda_-\, a_+ \quad \to\quad a_+(t)=a_+(0)e^{-i\lambda_- t}\nn\\
i\frac{\partial}{\partial t} a_-&=[a_-, H]=\lambda_+[a_-,a_+^\dagger]a_-=\lambda_+\, a_- \quad \to\quad a_-(t)=a_-(0)e^{-i\lambda_+ t}
\end{align}
so that, denoting by $U(t)$ the time evolution operator,
\begin{align}
U(-t)\>a_\pm U(t)&=a_\pm \>e^{-i\lambda_\mp t}\nn\\
U(-t)\>a_\pm^\dagger U(t)&=a_\pm ^\dagger\>e^{i\lambda_\pm t}
\end{align}
we find for the projectors
\begin{align}
U(-t)\>\vert \pm \rangle\langle \mp \vert\>&=U(-t) a_\pm^\dagger U(t)\> U(-t)\vert 0\rangle\langle \mp \vert=
e^{i\lambda_\pm t}\>\vert \pm \rangle\langle \mp \vert\nn\\
\vert \pm \rangle\langle \mp \vert\>U(-t)&=\vert \pm\rangle\langle 0\vert U(-t)\> U(t)a_\mp U(-t)=
\vert \pm \rangle\langle \mp \vert\>e^{i\lambda_\pm t}.
\label{time}
\end{align}

\vskip 10pt

We observe that quite naturally, a damping and anti-damping effect arises from the imaginary parts of the eigenvalues.
The two projectors have a time dependence factor $\exp(\mp\gamma t)$, with $\gamma>0$. However, the anti-damping term $\exp(+\gamma t)$
might give problems in the forward time evolution of zero-norm states. Actually, despite the
non-unitary representation of the time evolution operator, the scalar product is conserved since, using Eq.(\ref{time}),
\BE
\vert\pm\rangle_t=U(t) \vert\pm\rangle=e^{-i\lambda_\pm t}\vert\pm\rangle
\EE
and then
\begin{align}
_t\langle +\vert -\rangle_t&=\langle +\vert -\rangle\nn\\
_t\langle -\vert +\rangle_t&=\langle -\vert +\rangle
\end{align}
ensuring that all non-zero scalar products are invariant. Thus, in that sense, time evolution is unitary.

\vskip 10pt

Quite interestingly, it can be easily checked that the two-particle state 
$\vert +-\rangle=a_+^\dagger a_-^\dagger\vert 0\rangle$ has a positive norm $\langle +-\vert +-\rangle=1$ and is an eigenvector
of the Hamiltonian with a real eigenvalue $2\omega$ and no damping.  In a more realistic model 
this two-particle state could be interpreted as a glueball. Here, in this basic model, 
there is no explicit correlation between single-particle states and the two-particle state
appears as the product of single-particle states. In a more refined approach, it was argued that
i-particles, with complex conjugated energies, might give rise to well behaved two-particle propagators
and a real spectrum of physical vacuum excitations\cite{iparticle}.

\subsection{Propagator and complex poles}

For  a real field,  we would like to recover the principal part of the Minkowskian propagator by the anomalous set
of intermediate states that emerges when the positive- and negative-norm states are coupled in the non-perturbative limit. 
Because of the reported\cite{xigauge}
gauge-parameter independence of the principal part, we would like to focus on positive-norm intermediate states that might emerge in the
spectrum by the action of the pair of creation operators $a_\pm^\dagger$ for the zero-norm states. In the previous sections, the need for a pair of eigenvalues $E=\omega-i\gamma$ and $-E^*=-\omega-i\gamma$ emerged rather than a pair of complex conjugated
eigenvalues $\lambda_\pm$. Taking $E=\lambda_-$ we are tempted to assume that $-E^*=-\lambda_+$, which can be seen as the opposite of the energy of a particle. It is suggestive to interpret such state as the missing of a particle, as in the Dirac-sea language. The hypothesis is corroborated
by the observation that the vacuum might be defined modulo an arbitrary set of zero-norm states, as it happens in the Gupta-Bleuler formalism.
For instance, we can define a new vacuum $\vert \Omega\rangle$ as
\BE
\vert \Omega\rangle=\vert 0\rangle +c_-\vert -\rangle=\left[1+c_-a_-^\dagger\right]\vert 0\rangle
\EE
where $c_-$ is a constant, without affecting the norm $\langle \Omega\vert \Omega\rangle=\langle 0\vert 0\rangle$ and the expectation value of the observables, e.g.  $\langle \Omega\vert H\vert \Omega\rangle=\langle 0\vert H\vert 0\rangle=0$. The new vacuum is not annihilated by
$a_+$ which removes a particle yielding a physical (positive-norm) eigenstate
\BE
\vert \Psi_0\rangle=a_+\vert \Omega\rangle=c_-\vert 0\rangle
\EE 
with eigenvalue $E_0=0$. On the other hand, $a_+^\dagger$ adds a particle yielding another physical (positive-norm) state
\BE
\vert \Psi_2\rangle=a_+^\dagger\vert \Omega\rangle=\vert +\rangle+c_-\vert +,-\rangle.
\EE 
The original operators $a_0$, $a_1$ have non-zero matrix elements between the vacuum $\vert \Omega\rangle$ and the physical
intermediate states $\vert \Psi_0\rangle$, $\vert \Psi_2\rangle$ and could give rise to an anomalous principal part in the propagator.

However, the new vacuum $\vert \Omega\rangle$ acquires a time dependence and is not invariant by time evolution. 
In the Gupta-Bleuler formalism, the vacuum is not gauge invariant, since the added zero-norm states depend on the gauge. However, no measurable effect arises by such dependence.
Here, the new definition of vacuum would depend on time evolution which is supposed to be an other symmetry transformation for the physical vacuum.
In fact, denoting by $\vert \Omega^\pm\rangle$ the following definitions of vacuum
\BE
\vert \Omega^\pm\rangle=\vert 0\rangle +c_\pm\vert \pm\rangle=\left[1+c_\pm a_\pm^\dagger\right]\vert 0\rangle
\EE
we can write
\BE
\vert\Omega^\pm(t)\rangle=U(t)\,\vert \Omega^\pm\rangle=\left[1+c_\pm(t) a_\pm^\dagger\right]\vert 0\rangle
\label{Omegapm}
\EE
where
\BE
c_\pm(t)=c_\pm(0)\,e^{-i\lambda_\pm t}\sim e^{\pm\gamma t}
\label{cpmt}
\EE
yielding, asymptotically,
\begin{align}
\lim_{t\to +\infty}\vert\Omega^-(t)\rangle&=\vert 0\rangle\nn\\
\lim_{t\to -\infty}\vert\Omega^+(t)\rangle&=\vert 0\rangle.
\end{align}
Thus the states $\vert \Omega^+(t)\rangle$, $\vert \Omega^-(t)\rangle$ can be regarded as IN and OUT states, respectively, 
at a finite time when the interaction is still on
\begin{align}
\vert\Omega^-(t)\rangle&=U(t,\infty)\vert 0\rangle_{OUT}\nn\\
\vert\Omega^+(t)\rangle&=U(t,-\infty)\vert 0\rangle_{IN}.
\end{align}
Asymptotically, if the interaction is switched off adiabatically, the IN and OUT vacuum
tend to the same vacuum $\vert 0\rangle$. According, the vacuum-to-vacuum fluctuation amplitude reads
\BE
Z=_{\>\>OUT}\!\!\langle 0\vert  U(+\infty,-\infty)\vert 0\rangle_{IN}=\langle \Omega^-(t)\vert\Omega^+(t)\rangle
\EE
but does not depend on $t$ since 
\BE
c_-^*(t) c_+(t)=c_-^*(0) c_+(0)
\EE
so that time homogeneity is satisfied. More generally, inserting two fields $A(t_1)$, $A(t_2)$, at times $t_1>t_2$, 
the time ordered correlator is
\BE
\Delta(t_1,t_2)=_{\>\>OUT}\!\!\langle 0\vert A(t_1)A(t_2)\vert 0\rangle_{IN}
=\langle \Omega^-(t_1)\vert A\,U(t_1)U(-t_2)\,A\vert \Omega^+(t_2)\rangle
\EE
and by insertion of a set of positive-norm physical intermediate eigenstates $\{\vert n\rangle\}$, with energies $E_n$,
\begin{align}
\Delta(t_1,t_2)=&\sum_n\langle \Omega^-(t_1)\vert A\vert n\rangle e^{-iE_n(t_1-t_2)}\langle n\vert A \vert\Omega^+(t_2)\rangle=\nn\\
=&\sum_n\langle 0\vert A\vert n\rangle \langle n\vert A \vert 0\rangle\,e^{-iE_n(t_1-t_2)}
+\sum_n\langle -\vert A\vert n\rangle \langle n\vert A \vert +\rangle\,e^{-iE_n(t_1-t_2)} c_-^*(t_1)c_+(t_2)
\label{D12}
\end{align}
where we assumed that $\langle \pm\vert A\vert n\rangle=0$ if $\langle 0\vert A\vert n\rangle\not=0$ and vice versa. By Eq.(\ref{cpmt}), 
the last factor reads
\BE
c_-^*(t_1)c_+(t_2)=c_-^*(0)c_+(0)\,e^{i\lambda_+ (t_1-t_2)}
\EE
so that $\Delta(t_1,t_2)=\Delta(t_1-t_2)$ and, again, the homogeneity of time is satisfied.

Quite interestingly, if we take the real (physical) field $A=a_1+a_1^\dagger$
\BE
A=a_1+a_1^\dagger=\frac{1}{\sqrt{2}}\left[(a_+^\dagger+a_-^\dagger)e^{-i\theta}+(a_+ + a_-)e^{i\theta}\right]
\EE
and limit the sum to physical (positive-norm) intermediate eigenstates,  the correlator is
\BE
\Delta(t_1,t_2)=\sum_n\langle -\vert A\vert n\rangle \langle n\vert A \vert +\rangle\,e^{-i(E_n-\lambda_+)(t_1-t_2)} c_-^*(0)c_+(0)+\dots
\EE
where the dots refer to a regular part which arises from real energies in the first sum of Eq.(\ref{D12}).
The remaining sum over $n$ can only include the positive-norm eigenstates $\vert 0\rangle$ and $\vert +,-\rangle$, with eigenvalues $0$ and $2\omega$, respectively. In fact, acting with $A$ over $\vert\pm\rangle$ can only give the further states 
$\vert ++\rangle$ and $\vert --\rangle$ which have zero norm.
The matrix elements read
\begin{align}
\langle 0\vert A \vert +\rangle&=\langle 0\vert A a_+^\dagger\vert 0\rangle=\frac{e^{i\theta}}{\sqrt{2}}\langle 0\vert a_- a_+^\dagger\vert 0\rangle
=\frac{e^{i\theta}}{\sqrt{2}}\nn\\
\langle 0\vert A \vert -\rangle&=\langle 0\vert A a_-^\dagger\vert 0\rangle=\frac{e^{i\theta}}{\sqrt{2}}\langle 0\vert a_+ a_-^\dagger\vert 0\rangle
=\frac{e^{i\theta}}{\sqrt{2}}\nn\\
\langle +,-\vert A \vert +\rangle&=\frac{e^{-i\theta}}{\sqrt{2}}\langle +,-\vert a_-^\dagger a_+^\dagger\vert 0\rangle
=\frac{e^{-i\theta}}{\sqrt{2}}\langle +,-\vert +,-\rangle
=\frac{e^{-i\theta}}{\sqrt{2}}\nn\\
\langle +,-\vert A \vert -\rangle&=\frac{e^{-i\theta}}{\sqrt{2}}\langle +,-\vert a_+^\dagger a_-^\dagger\vert 0\rangle
=\frac{e^{-i\theta}}{\sqrt{2}}\langle +,-\vert +,-\rangle
=\frac{e^{-i\theta}}{\sqrt{2}}.
\end{align}
Assuming that  $c_+(0)=c_-(0)=c(0)$ in Eq.(\ref{Omegapm}),  the time ordered correlator, for $t_1>t_2$, 
can be written as
\BE
\Delta(t_1-t_2)=\frac{\vert c(0)\vert^2}{\sqrt{2}} \left[e^{i \lambda_+(t_1-t_2)} + e^{-i \lambda_-(t_1-t_2)}\right]
\EE
with frequencies $-\lambda_+=-\omega-i\gamma$ and $\lambda_-=\omega-i\gamma$. 
The correlator has precisely the same structure of the effective propagator in Eq.(\ref{tprop}), containing the
same pair of complex frequencies  which led to the Minkowskian principal part of Eq.(\ref{Minkstruct})
and a positive weight emerging from the positive-norm states. 
The result is consistent with the structure of the effective gluon propagator in Minkowski space, as
described in Sec.~IV.

In Yang-Mills theory, a non-perturbative composite vacuum like that might  be related to the existence 
of a condensate $\langle A^2\rangle\not=0$ containing zero-norm states. Thus, we can speculate that a similar
mechanism might link the principal part of the gluon propagator with physical excited states,  
glueballs, produced by the binding of a single-particle  zero-norm excited state with a second zero-norm state which was
in the condensate. Because of the interaction with the condensate, that state would be damped and would appear as a
confined quasiparticle. On the other hand, the state $\vert +,-\rangle$ would also appear as a two-particle excitation
of the vacuum with an energy  $\delta E=E+E^\star=2\Rerm{M}\approx 2(0.581)$ GeV, 
according to the data of Ref.\cite{xigauge}. That energy would be compatible
with a glueball resonance which would overlap with light mesons.

\section{Discussion}

We have shown that the analytic continuation of the anomalous part of the gluon propagator leads to the definition
of an effective propagator in Minkowski space, which seems to be directly related to the eigenvalue spectrum
of the Hamiltonian. The change of sign in the anomalous part, containing the ``wrong'' pole, provides a
{\it physical} function which has the same identical structure which would arise from first principles if
complex eigenvalues were present in the spectrum. Moreover, the principal part of the effective propagator 
becomes imaginary in Minkowski space and defines a real spectral density which provides a generalized
K\"all\'en-Lehmann representation, including the principal part which had to be added by hand in the standard
formulation\cite{dispersion}. We argue that the modified spectral representation might
reconcile some inconsistencies which emerge in the spectral Schwinger-Dyson formalism\cite{spectral} 
when the usual definition of the spectral function is used\cite{wink}.

It is remarkable that the definition of the anomalous spectral density and its direct link to the eigenvalues can
only emerge because of the change of sign which is found going from the Euclidean to Minkowski space
through the clock-wise Wick rotation. Of course, the present analysis does not explain 
the origin and nature of the very peculiar pair of complex energies, with a negative imaginary part, which must be selected
for convergence reasons. But their existence is predicted by
the structure of the effective propagator and seems to be the only way to give a physical interpretation to the
gluon correlator from first principles.

While the existence of complex poles has been reported by many different approaches, ranging from
the Gribov-Zwanziger effective model\cite{GZ} 
to one-loop calculations by the screend expansion\cite{ptqcd,analyt,xigauge}, 
to truncations of Schwinger-Dyson equations 
in the complex plane\cite{fischer},
and has been even proven formally in loop expansions\cite{kondo18}, the genuine nature and role of these complex
poles is currently under debate.  The reported gauge invariance of poles and residues\cite{xigauge} seems to favor
a genuine physical role.

In Ref.\cite{kondo21}, from a formal point of view, taking for granted
the usual analytic continuation in time, it is shown that a gluon propagator with complex poles cannot even be  
defined in the Minkowski direct space and
the related gluon degrees of freedom should be regarded as unphysical. That is quite disappointing, since
complex poles have bee found even in quark propagators\cite{quark}.

We take the opposite view that a confined gluon is still a physical object and that the gluon correlator must be defined
somehow in the Minkowski space. From that physical argument and from the necessity of an anomalous clock-wise rotation,
a finite result can be found only if that selected pair of complex energies is taken, with negative imaginary part.
The nature of these energies remains obscure. 

Complex eigenvalues of an Hermitian Hamiltonian emerge in
a pseudo-Euclidean space where negative-norm states are present.
The complex energies are associated with zero-norm states which are usually regarded as unphysical\cite{kondo21}.
Thus, the  intermediate states which give rise to the main contribution to the gluon propagator might be
just unphysical.

However, an untrivial vacuum structure might contain a superposition of zero-norm states like a sort of condensate. 
In that case,
as suggested by a toy model at the end of  the previous section, the intermediate states might be
physical states, with a positive norm, and yet be characterized by the occurrence of complex frequencies
in the real-time propagator.
In simple words, an excited zero-norm state might be correlated with an other zero-norm state which already
is in the vacuum, but has an opposite imaginary part of the energy,  yielding a {\it physical} pair with a real total
energy. 
The real-time and imaginary-time correlators predict that the quasiparticle would be anyway
damped at the scale
of 1 fm, providing a dynamical mechanism of confinement. Thus, in that case, behind the complex frequencies 
there would be physical states which would {\it appear} as damped single-particle  confined quasiparticles.
On the other hand,  the two-particle states would also appear as glueball excitations
of the vacuum with an energy  $\delta E=E+E^\star=2\Rerm{M}\approx 1$ GeV and
could easily mix with known meson resonances. 

In any case, any further analysis should take in due account the existence of an untrivial, anomalous 
analytic continuation when going to Minkowski space, with important consequences on the physical
interpretation of the theory.

\acknowledgments

This research was supported in part by the INFN-SIM national project and by the ``Linea di intervento 2'' 
for HQCDyn at DFA-Unict.

\end{document}